\documentclass[
    aps,
    prd,
    twocolumn,
    superscriptaddress,
    nofootinbib,
    amsmath,
    amssymb,
]{revtex4-2}

\usepackage[utf8]{inputenc}
\usepackage[T1]{fontenc}          
\usepackage{lmodern} 
\usepackage[british]{babel}       
\usepackage{microtype}            

\usepackage{aas_macros}  
\usepackage{graphicx}  
\usepackage{dcolumn}  
\usepackage{bm}  
\usepackage{url}
\usepackage{subcaption}
\usepackage{amsfonts}
\usepackage{array}
\usepackage{multirow}
\usepackage{xcolor}
\usepackage{ragged2e}  
\usepackage[pdftex, colorlinks=true, allcolors=blue]{hyperref}

\makeatletter
\long\def\@makecaption#1#2{%
  \vskip\abovecaptionskip
  \sbox\@tempboxa{\small#1. #2}%
  \ifdim \wd\@tempboxa > \hsize
    {\small\justifying#1. #2\par}
  \else
    \global\@minipagefalse
    \hb@xt@\hsize{\hfil\box\@tempboxa\hfil}%
  \fi
  \vskip\belowcaptionskip}
\makeatother

\newcommand*{\valencia}{Departamento de Astronom\'{\i}a y Astrof\'{\i}sica, Universitat de Val\`encia, Dr.~Moliner 50,  46100 Burjassot (Val\`encia), Spain.}

\newcommand*{\OAUV}{Observatori Astron\`omic, Universitat de Val\`encia, Catedr\'atico Jos\'e Beltr\'an 2, 46980 Paterna (Val\`encia), Spain}

\begin{document}

\title{Classification of the equation of state of neutron stars via sparse dictionary learning}

\author{Miquel Llorens-Monteagudo} \email{Contact author: miquel.llorens@uv.es} \affiliation{\valencia}
\author{Alejandro Torres-Forn\'e} \affiliation{\valencia} \affiliation{\OAUV}
\author{Jos\'e A.~Font} \affiliation{\valencia} \affiliation{\OAUV}

\date{\today}

\begin{abstract}
    The post-merger phase of binary neutron star (BNS) mergers encodes valuable information about the equation of state (EOS) of supranuclear matter. Extracting this information from the analysis of the post-merger waveforms remains challenging due to the high-frequency limitations of current detectors. Future third-generation observatories, such as the Einstein Telescope (ET) and NEMO, will have the sensitivity required to resolve post-merger signals with high fidelity. In this work, we apply \textsc{clawdia}, our recently developed sparse dictionary learning (SDL) framework,  to classify different EOS models using only the post-merger gravitational-wave emission of simulated BNS mergers available in the \textsc{CoRe} database. Our dataset comprises five EOS models representative of a broad range of neutron star properties. The SDL framework is optimised under realistic detection conditions by injecting signals into simulated noise matching the sensitivity curves of ET and NEMO. Our results show that classification is primarily driven by the dominant post-merger frequency, $f_2$, which encodes EOS-dependent information. At a modest signal-to-noise ratio of 5, our method achieves $F_1$ scores of $0.76$ for ET and $0.70$ for NEMO, with performance improving for higher signal-to-noise ratios. The reliability and generalisation capabilities of the model are assessed with additional tests, including the classification of an EOS not included in the training dataset and the analysis of detector-specific biases.
\end{abstract}


\maketitle

\section{Introduction}\label{sec:introduction}

The advent of gravitational-wave (GW) astronomy has opened a powerful observational window into stellar-origin compact objects, enabling us to test and refine current theoretical models of neutron stars and black holes~\cite{GWTC-1,GWTC-2,2023PhRvX..13d1039A,GWTC-4}. In particular, the seminal observation of GWs from binary neutron star (BNS) merger GW170817~\citep{GW170817,GBM:2017lvd}, placed constraints on the equation of state (EOS) and radius of neutron stars by measuring the tidal deformability from the analysis of the inspiral waveform. Those properties were inferred through Bayesian statistical methods by matching the collected data with predicted waveforms from general relativity~\cite{2018PhRvL.121p1101A,2019PhRvX...9a1001A}. 

The extraction of neutron star information from the inspiral can be complemented by the analysis of the post-merger signal. While searches for such a signal were conducted for GW170817, no detection was reported~\cite{GW170817-post-merger}. This was not unexpected as the frequency of the post-merger signal is above the sensitivity limit of the LIGO-Virgo-KAGRA (LVK) detector network at high frequencies. Progress on our understanding of the post-merger waveform entirely relies on numerical relativity (NR) simulations~\cite{Bernuzzi:2020}. Those have revealed a rich GW phenomenology, with spectral features dominated by distinctive peaks associated with specific oscillation modes of the remnant (see e.g.~\cite{Topolski:2023} and references therein for a recent study on this topic). Over the years there has been increased interest in identifying quasi-universal (EOS-insensitive) relations between oscillation frequencies (spectral peaks) and neutron star properties (e.g.~mass, radius, tidal deformability). Some features, such as the main post-merger quadrupolar frequency $f_2$, have been studied extensively whereas other features like secondary peaks or late-time inertial modes triggered by convection, have gained more attention recently~\cite{Bauswein:2012,Read:2013,Takami:2015,Bauswein:2015,DePietri:2018,DePietri:2020,Topolski:2023}. 

Numerically generated waveforms of BNS mergers constitute invaluable datasets to perform parameter inference tasks of the source properties. Recent examples include the reconstruction of post-merger waveforms injected in simulated detector noise to constrain the EOS of neutron star matter~\cite{Chatziioannou:2017,Miquel:2023,Miquel:2025}. The availability of simulations of BNS mergers, however, remains limited due to their high computational cost and large parameter space. This limitation underscores the importance of developing alternative approaches for parameter estimation, capable of generalising from limited datasets. While machine-learning techniques based on convolutional or residual neural networks have shown a great potential for waveform classification and parameter estimation (see~\cite{Cuoco:2021,Cuoco:2025} and references therein), they often face limitations when training data is scarce, further exacerbated by the complexity of the background noise inherent to GW detectors. In such cases Sparse Dictionary Learning (SDL) algorithms offer a promising alternative~\cite{Elad:2006}. 

SDL achieves a sparse representation of GW data through the linear combination of basic elements of the GW signals making up a dictionary, dubbed as `atoms'. SDL has emerged as a compelling alternative technique to traditional signal representation approaches and very efficient methods have been devised to solve the optimisation problem inherent to learning dictionaries~\cite{Mairal:2012}. By learning a compact, data-driven representation of the waveform space, SDL algorithms can enhance the generalisation capability, providing more robust behaviour in the presence of high-dimensional noise within GW detectors\footnote{The term ``high-dimensional noise'' refers to the complexity of the background noise in GW detectors, which arises from multiple independent and interdependent sources, such as seismic activity, thermal vibrations, quantum noise, and anthropogenic disturbances. This noise is further characterized by its non-stationary nature, as evidenced by the gradual variation of the detector's PSD sensitivity over time.}.
Recently, applications of SDL have been achieved in the field of GW data analysis. Those range from the removal of instrumental noise from GW detectors to the reconstruction of signals in different astrophysical contexts~\cite{Torres-Forne:2016,Llorens-Monteagudo:2019,Torres-Forne:2020,Ainara:2022,Badger:2023,Powell:2024,Badger:2024a,Badger:2024b}.

In this paper we assess the performance of SDL algorithms to classify the EOS of neutron stars. To do so we employ our own computational framework called \textsc{clawdia}~\cite{CLAWDIA}, a modular Python package designed to bring together SDL-based methods for GW data analysis. \textsc{clawdia} provides an interface that simplifies the application of SDL techniques, and currently includes a modular pipeline for signal classification which integrates typical stages such as pre-processing and denoising. Our study employs data from BNS merger simulations corresponding to various EOS, analysing exclusively the post-merger GW emission. Specifically, we use the \textsc{CoRe} database~\cite{CoRe_2023}, a publicly accessible repository containing an extensive collection of NR simulations of BNS mergers. Waveforms from this database are injected into simulated noise mimicking the expected sensitivities of the third-generation detectors Einstein Telescope (ET)~\cite{ET:2010} and NEMO~\cite{NEMO}. This choice is motivated by the fact that typical post-merger GW frequencies are above 1kHz, which greatly challenges detection with present-day interferometers. The injected signals are whitened using the corresponding design power spectral density (PSD) of the detectors. For each EOS, noise-free signals are used to initialize and train the dictionaries intended for denoising and classification, while the noise-injected signals are used to optimise the parameters and validate the model. 

As we show below, the performance of the \textsc{clawdia} pipeline is strongly conditioned by the GW spectral features, especially those associated with the dominant post-merger quadrupolar mode. In particular, the relative proximity of the spectral peaks for certain classes of EOS represents a challenge for the classification process. However, our SDL algorithm successfully identifies the correct EOS even for low values of the signal-to-noise ratio (SNR). We estimate that the minimum SNR required for reliable EOS classification is about 5, with the method's performance stabilising at higher SNR values. These observations demonstrate the robustness of our pipeline and its potential applicability in realistic observation scenarios.

The organization of the paper is as follows: Section~\ref{sec:dataset} describes the dataset used in this study, including the selection of EOS models and the characteristics of the NR simulations from the \textsc{CoRe} database. Section~\ref{sec:classification-model} describes the classification model, outlining the mathematical framework of SDL and the pipeline architecture. Section~\ref{sec:results} presents the results of our analysis, including pipeline optimisation, performance across varying SNR levels, and classification of signals from an unseen EOS (meaning an EOS absent from both training and pipeline optimisation). Finally, Section~\ref{sec:discussion} discusses the implications of our findings, compares our results with related work, and outlines potential directions for future research.

\section{THE DATASET}\label{sec:dataset}

At the time of its second release, the \textsc{CoRe} database contains 590 individual simulations, corresponding to 254 distinct BNS configurations, and spans a total of 18 different EOS models for the neutron star matter~\cite{CoRe_2023}. The simulations explore a wide range of parameters, including (a) total binary masses ranging from 2.4~$M_\odot$ to about 3.4~$M_\odot$, with mass ratios up to $q = 2.1$, (b) EOS stiffness, reflected in the dimensionless tidal polarizability parameters $\Lambda_1$ and $\Lambda_2$, which strongly influence the waveform evolution, and (c) spin-orbit interaction, with dimensionless spin components up to $\chi_z = 0.5$. For each simulation, \textsc{CoRe} provides both detailed metadata and the actual GW data. The GW outputs include the strain polarisations, $Rh_+$ and $Rh_\times$, where $R$ is the extraction radius, as well as the Weyl curvature multipoles, $\Psi_4$, computed, in most cases, up to $(\ell, m) = (4, 4)$. Time and distance are rescaled in terms of the binary mass, $M = m_1 + m_2$, where $m_{1,2}$ are the gravitational masses of the individual stars. To access the \textsc{CoRe} database, we use the \textsc{Watpy} \cite{Watpy_2024} Python package, which offers an efficient interface for retrieving both metadata and waveform data. It allows users to easily filter NR simulations by their parameters, bulk download selected models, and perform various other operations.

\subsection{EOS selection}

\begin{figure}
    \centering
    \includegraphics[width=0.95\columnwidth]{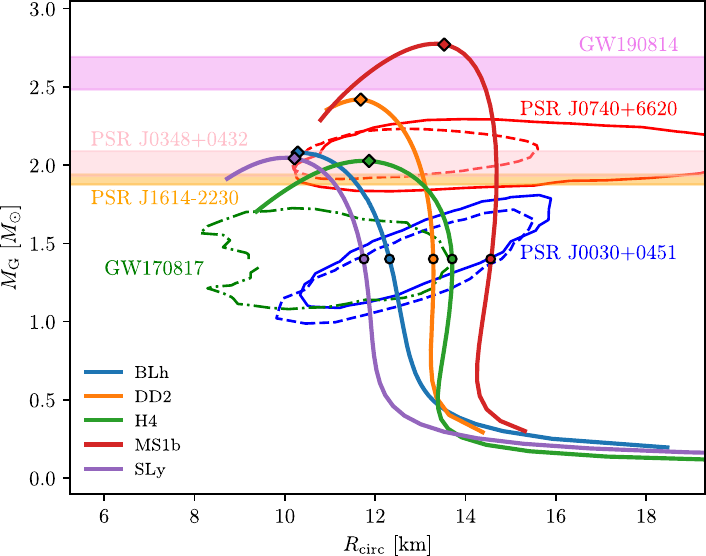}
    \caption{
        Mass-radius sequences of the EOS included in the dataset. Diamond-shaped markers correspond to the maximum mass for each EOS, whereas the circle-shaped ones show the radius $R_{1.4}$ for a 1.4~$M_\odot$ star. The color bands show the current mass constraints of GW190814 \cite{GW190814}, PSR~J0348+0432 \cite{Antoniadis:2013}, and PSR~J1614-2230 \cite{Arzoumanian:2018}. The contour lines show the combined constraints of GW170817 \cite{GW170817}, PSR~J0740+6620 \cite{Riley:2021,Miller:2021}, and PSR~J0030+0451 \cite{Riley:2019,Miller:2019}.
    }
    \label{fig:bns-dataset-mass-radius}
\end{figure}

The set of EOS included in the dataset encompasses a range of theoretical models for neutron star matter, each differing in their physical assumptions, predicted properties, and compatibility with observational data. The selection criteria are based on two key requirements: continued relevance within the scientific community and sufficient representation in the \textsc{CoRe} database to enable the division of the dataset into standard training and test subsets. Based on these criteria, we selected five EOS, which define the different classes of our model: SLy, MS1b, H4, BLh, and DD2. Table~\ref{tab:eos-physical-properties} provides a summary of key parameters, and the (gravitational) mass-radius relations are illustrated in Figure~\ref{fig:bns-dataset-mass-radius}, with the maximum masses and radii for a 1.4~$M_\odot$ neutron star highlighted. The agreement of the set of EOS with observational data, including GW constraints from GW170817 and NICER observations of pulsars like PSR J0030+0451 \cite{Miller:2019} and PSR J0740+6620 \cite{Riley:2021,Biswas:2021}, varies among models. In general, stiffer models like MS1b and H4 face greater challenges to accommodate observational constraints compared to softer models such as SLy and BLh. However, we keep the latter two EOSs in our dataset for completeness, since our main purpose is to estimate the classification efficiency of our SDL method for a broad range of physical models.

\begin{figure*}
    \centering
    \begin{subfigure}[b]{0.48\linewidth}
        \caption{MS1b}\label{fig:EoS_signals:ms1b}
        \includegraphics[width=\linewidth]{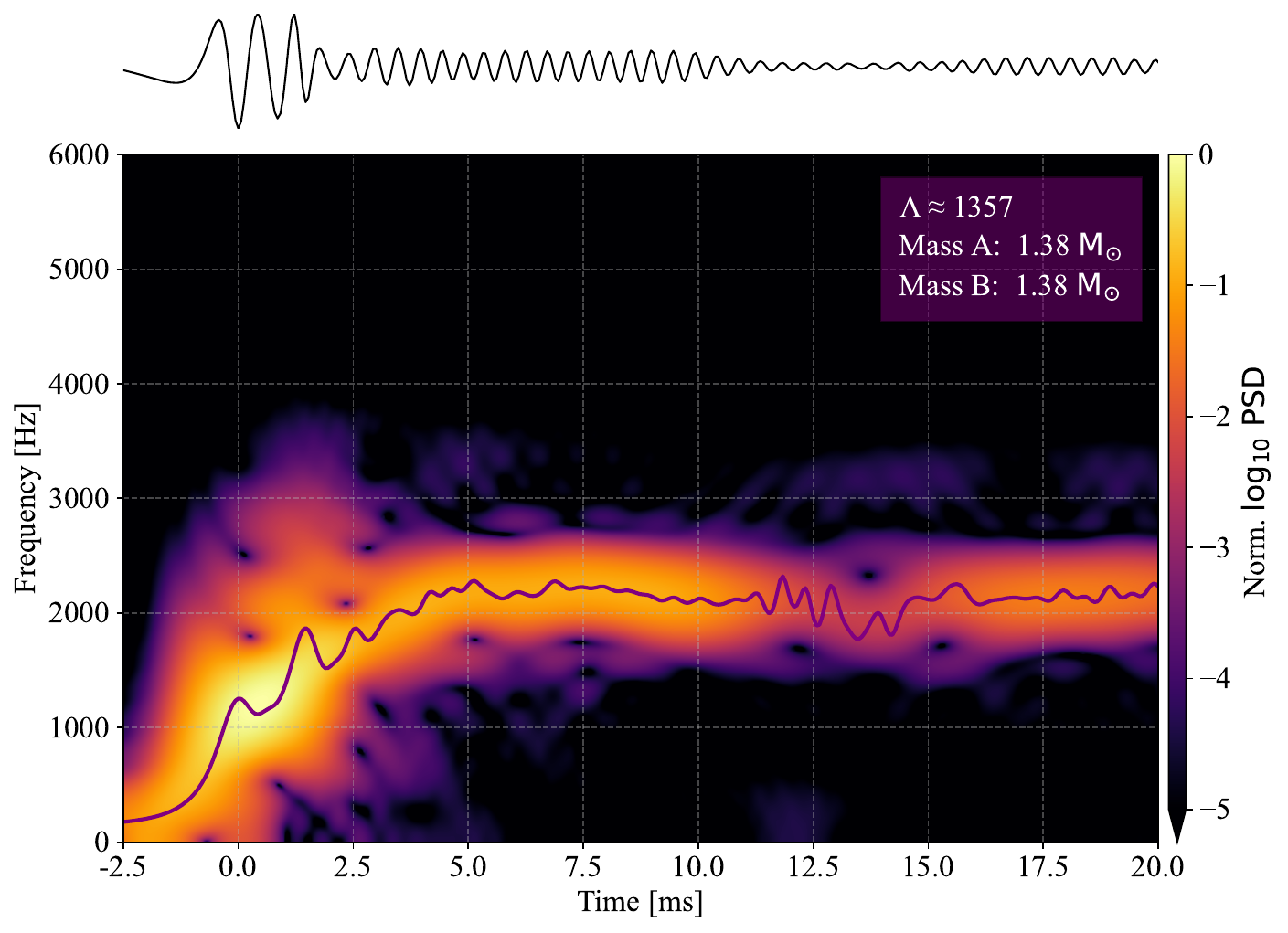}
    \end{subfigure} \hfill
    \begin{subfigure}[b]{0.48\linewidth}
        \caption{H4}\label{fig:EoS_signals:h4}
        \includegraphics[width=\linewidth]{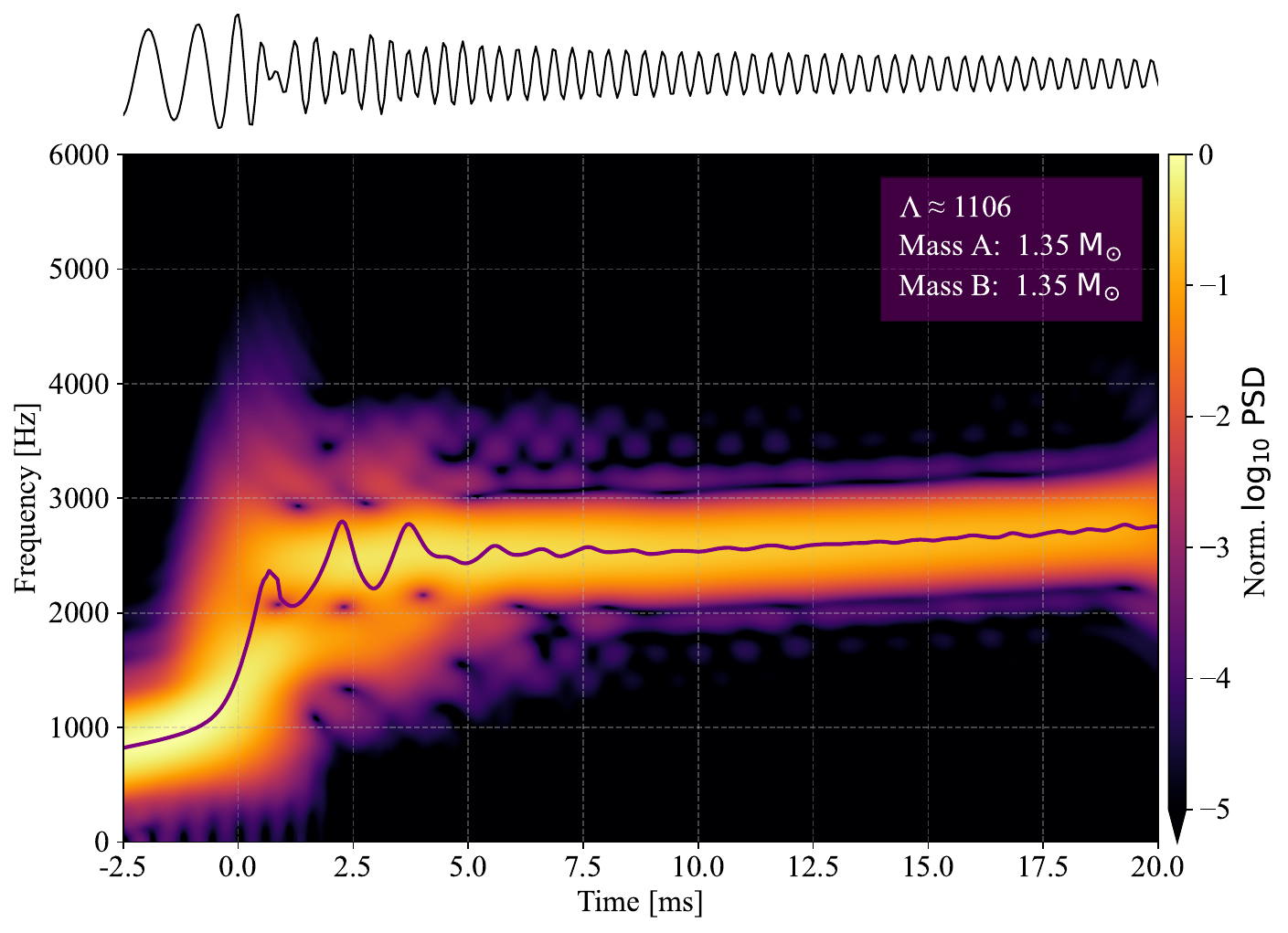}
    \end{subfigure}

    \vspace{1em}

    \begin{subfigure}[b]{0.48\linewidth}
        \caption{DD2}\label{fig:EoS_signals:dd2}
        \includegraphics[width=\linewidth]{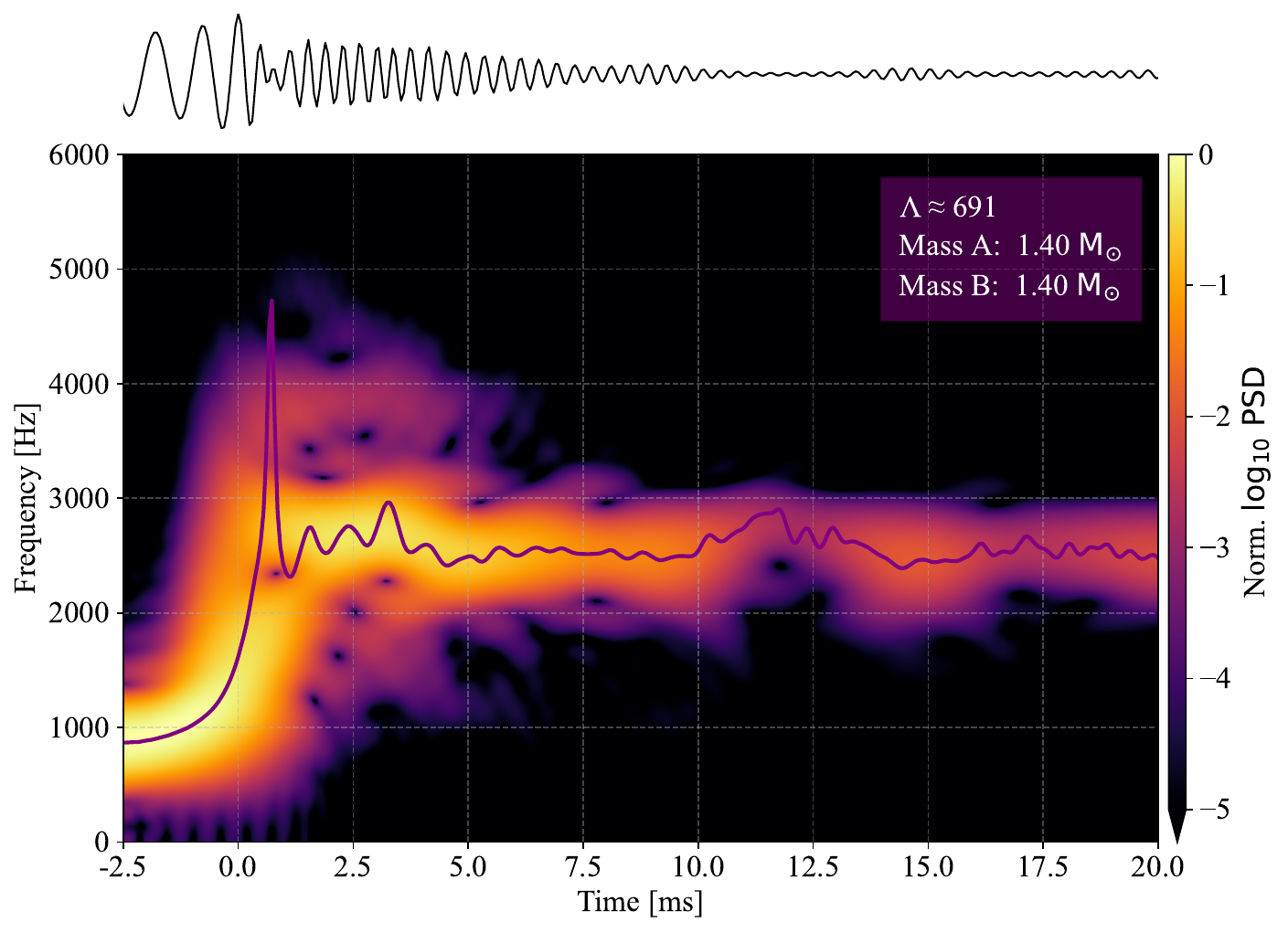}
    \end{subfigure} \hfill
    \begin{subfigure}[b]{0.48\linewidth}
        \caption{BLh}\label{fig:EoS_signals:blh}
        \includegraphics[width=\linewidth]{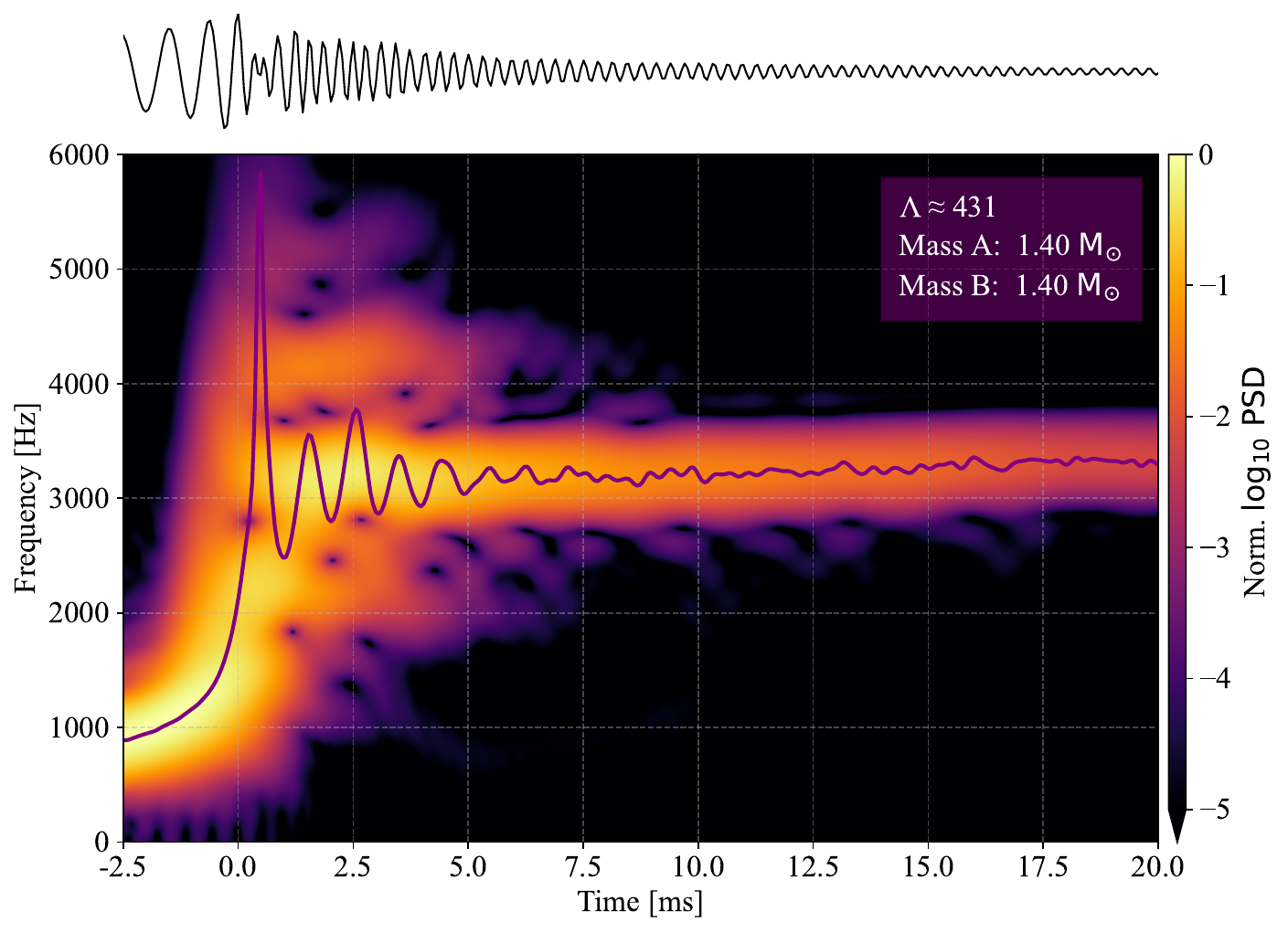}
    \end{subfigure}

    \vspace{1em}

    \begin{subfigure}[b]{0.48\linewidth}
        \caption{SLy}\label{fig:EoS_signals:sly}
        \includegraphics[width=\linewidth]{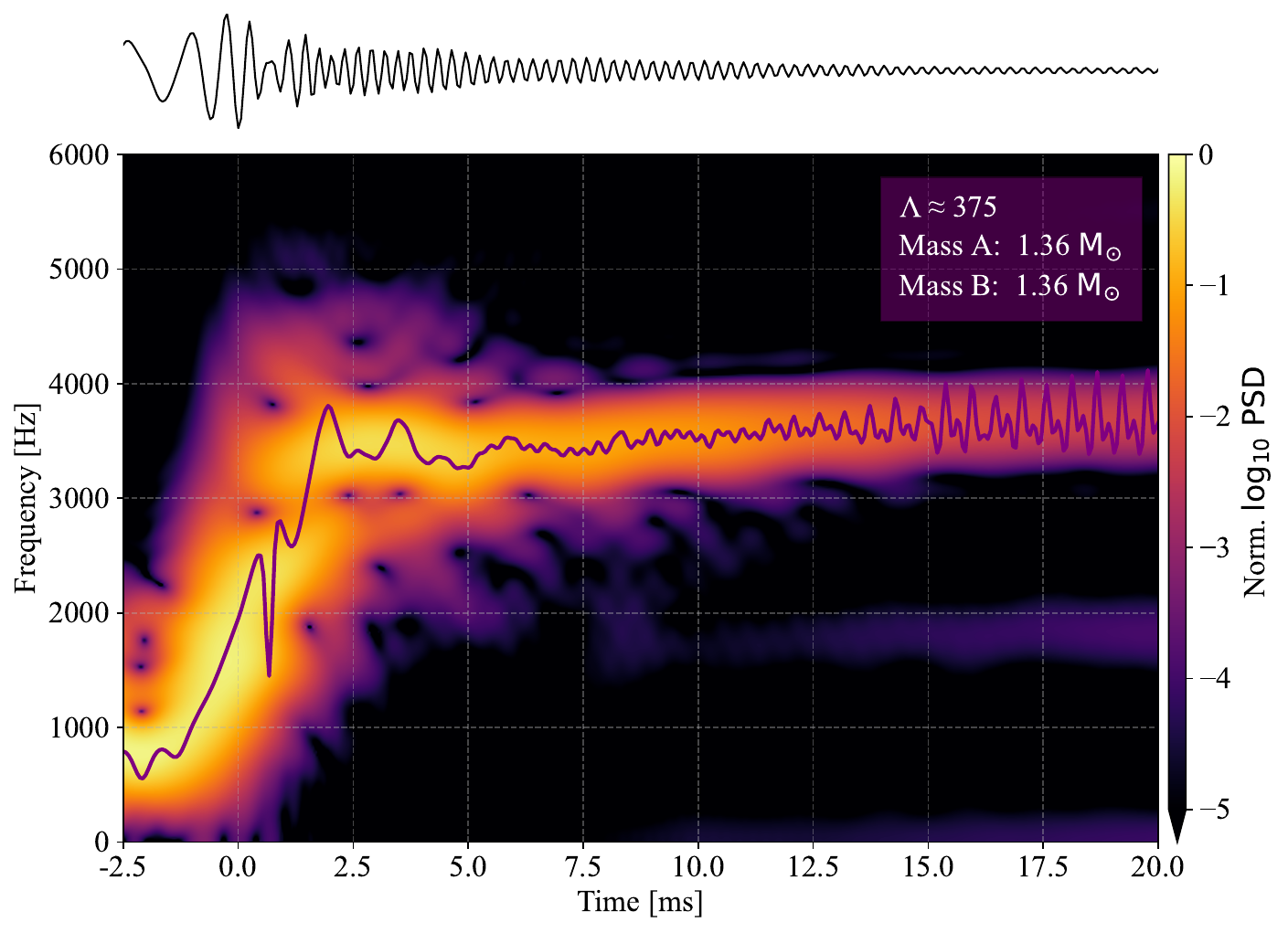}
    \end{subfigure}
    
    \caption{
        Gravitational wave spectrograms from NR simulations of BNS mergers with different EOS models. Each panel shows the normalized power spectral density (PSD) on a logarithmic scale, covering the late inspiral, merger (centered at $t=0$), and post-merger phase. Solid curves represent instantaneous frequencies. $\Lambda$ denotes the dimensionless tidal deformability, while masses A and B correspond to individual neutron stars. The time-domain GW signals are overlaid at the top of each panel.
    }
    \label{fig:EoS_signals}
\end{figure*}

For each model, we display in Figure~\ref{fig:EoS_signals} a representative simulation of a BNS merger with an initial quasi-circular orbit, equal-mass ratio, and initial gravitational masses as close to $1.4~M_\odot$ as available in the database. The GWs in the panels of  Figure~\ref{fig:EoS_signals} are represented in three formats. At the top, we display the original time-domain waveform, as obtained from \textsc{CoRe}. In the main plot, we show both the instantaneous frequency over time (purple line) and its spectrogram, for it is a powerful tool for visualizing spectral features, specially during the transient post-merger phase, which may be obscured by the dominant frequency peak when computing the full PSD \cite{Rezzolla:2016}.

\begin{table}
    \caption{ 
        Overview of the properties of each EOS in the dataset, including whether it incorporates exotic phases of matter (e.g., hyperons or deconfined quarks) and if it is based on relativistic theory. The table also provides the stiffness classification, maximum supported mass $M_\text{max}$, radius $R_{1.4}$, and dimensionless tidal deformability $\Lambda_{1.4}$ for a neutron star with mass 1.4~M$_\odot$.
    }
    \label{tab:eos-physical-properties}
    \begin{ruledtabular}
    \begin{tabular}{lcccccc}
        EOS & Exotic & GR & Stiffness & $M_\text{max}$ (M$_\odot$) & $R_{1.4}$ (km) & $\Lambda_{1.4}$ \\ 
        \hline
        \noalign{\vskip 3pt}  
        MS1b  & No  & Yes  & Very Stiff & 2.77 & 14.5 & 1220 \\
        H4    & Yes & Yes  & Mod. Stiff & 2.03 & 13.8 & 900 \\
        DD2   & No  & Yes  & Mod. Stiff & 2.42 & 13.2 & 674 \\
        BLh   & No  & No   & Mod. Stiff & 2.10 & 12.5 & 510 \\
        SLy   & No  & No   & Mod. Soft  & 2.05 & 11.7 & 300 \\
    \end{tabular}
    \end{ruledtabular}
\end{table}

\subsubsection{MS1b EOS}

The MS1b EOS, based on relativistic mean-field theory with nonlinear meson interactions, is the stiffest model in our dataset and includes a first-order (Van der Waals) phase transition from hadronic to quark matter. It predicts a maximum mass of 2.78~M$_\odot$, a 14.5~km radius for a 1.4~M$_\odot$ star, and a high tidal deformability $\Lambda_{1.4} \approx 1220$ \cite{Biryukov:2017,Baiotti:2019}. Constraints from GW170817 \cite{GW170817} and NICER observations \cite{Riley:2021} challenge this EOS due to evidence for smaller radii and lower deformabilities (see Figure~\ref{fig:bns-dataset-mass-radius}). A recent study on inflationary attractors \cite{Odintsov:2023} further places MS1b’s radii near current causal limits. Despite this, it remains widely used in \textsc{CoRe} and serves as an extreme case in the mass–radius space. An example of GW evolution from a BNS merger using MS1b is shown in Figure~\ref{fig:EoS_signals:ms1b}. The merger frequency, $f_\text{mer} \approx 1.25~\text{kHz}$, is relatively low but consistent with this stiff EOS. The main post-merger peak at $f_2 \approx 2.1~\text{kHz}$ exhibits a low-frequency modulation with core bounces at roughly 2.5 and 12.5~ms, matching the large tidal deformability ($\Lambda \approx 1357$) that sustains a long-lived remnant.

\subsubsection{H4 EOS}

The H4 EOS, developed within relativistic mean-field theory, incorporates hyperons to model high-density matter \cite{Lackey:2006}, which leads to a softening of the EOS at high densities, though the overall stiffness remains moderate. 
This EOS predicts a maximum mass of  $\approx 2.03$~M$_\odot$, a radius of 13.8~km for a 1.4~M$_\odot$ neutron star \cite{Lackey:2006}, and a moderately high tidal deformability ($\Lambda_{1.4} \approx 900$). The H4 EOS is affected by the uncertainties surrounding hyperon interactions and by recent observational mass-radius constraints from NICER \cite{Riley:2021}. Figure~\ref{fig:EoS_signals:h4} illustrates the GW evolution from a BNS merger simulated using this EOS. The moderately high stiffness of the H4 EOS results in a low merger frequency ($f_\text{mer} \approx 1.54~\text{kHz}$) and post-merger peak ($f_2 \approx 2.57~\text{kHz}$). Notably, the widening of the spectral energy within the first 5~ms after merger is common in cases where a hypermassive neutron star (HMNS) forms, which in this case is supported by the high maximum mass of the H4 model.
The remnant's lifetime is shorter than in the previous example, and the dominant frequency $f_2$ shifts towards higher values. This reflects the gradual loss of angular momentum through GW emission, leading to reduced rotational support and a higher central density (until collapse, not included in the plot).

\subsubsection{DD2 EOS}

The DD2 EOS is a relativistic mean-field model with density-dependent couplings that accounts for medium effects on nucleons and clusters at low densities \cite{Typel:2010}. It predicts a maximum mass of 2.42~M$_\odot$, a radius of 13.2~km for a 1.4~M$_\odot$ neutron star, and a tidal deformability of $\Lambda_{1.4} \approx 674$ \cite{Zhu:2018}. DD2 provides a smooth transition from low-density to high-density regimes, making it applicable across a wide range of conditions. However, its predicted pressures at high densities exceed those suggested by experimental data from heavy-ion collisions \cite{Danielewicz:2002,Nedora:2021}. Despite this, DD2's predictions for radii and tidal deformabilities remain consistent with the GW170817 observation, and its maximum mass exceeding 2~M$_\odot$ supports the existence of massive neutron stars, such as MSP~J0740+6620 \cite{Cromartie:2020} (see Figure~\ref{fig:bns-dataset-mass-radius}). Figure~\ref{fig:EoS_signals:dd2} shows the GW signal of a BNS merger simulation with the DD2 EOS. The frequency at merger is $f_\text{mer} \approx 1.62~\text{kHz}$ and the HMNS survives throughout the simulation ($\gtrsim 22~\text{ms}$), supported by differential rotation. The remnant undergoes several contractions and oscillations during post-merger, apparent in the GW signal as low-frequency modulations and a less steady main frequency at $f_2 \approx 2.5~\text{kHz}$. 

\subsubsection{BLh EOS}

The BLh EOS is a microscopic model derived from Brueckner–Bethe–Goldstone many-body theory \cite{Bombaci:2018,Logoteta:2021}. While it provides a good description of nuclear matter at finite temperature, it does not include exotic phases such as hyperons or deconfined quarks. BLh is a moderately soft EOS, predicting a maximum neutron star mass of $\approx 2.08$~M$_\odot$, a radius of 12.5~km for a 1.4~M$_\odot$ neutron star, and a tidal deformability of $\Lambda_{1.4} \approx 510$ \cite{Bernuzzi:2020,Cusinato:2022}. BLh has produced predictions consistent with the blue kilonova component observed in the electromagnetic counterpart of GW170817, AT\,2017gfo \cite{Nedora:2021}, making it a promising candidate for future studies.
Figure~\ref{fig:EoS_signals:blh} displays the GW signal of a BNS merger simulations conducted with the BLh EOS, with 
$\Lambda \approx 431$. The neutron stars merge at $f_\text{mer} \approx 2.12~\text{kHz}$, forming a stable HMNS that survives for the duration of the simulation ($\gtrsim 39~\text{ms}$). The post-merger phase exhibits a sustained dominant frequency of $f_2 \approx 3.2~\text{kHz}$.

\subsubsection{SLy EOS}

The SLy EOS, the final one in our dataset, is a non-relativistic model derived from Skyrme Lyon effective nuclear interactions, particularly suited for neutron-rich matter \cite{Douchin:2001}. It is the softest EOS in our dataset, with a maximum mass of 2.05~M$_\odot$, a relatively small radius of 11.7~km for a 1.4~M$_\odot$ neutron star, and a low tidal deformability ($\Lambda_{1.4} \approx 300$) \cite{Hinderer:2010,Biryukov:2017}. SLy includes a weak first-order phase transition between the crust and core, but does not account for exotic phases such as hyperons or quark deconfinement. While the non-relativistic nature of this EOS raises concerns about its validity at extreme densities, it shows marginal agreement with the lower boundary of the allowed $\Lambda$ region for GW170817 \cite{Radice:2018}. Figure~\ref{fig:EoS_signals:sly} displays the GW signal representative of the SLy EOS for a BNS merger simulation, with $\Lambda \approx 375$. The two stars merge at $f_\text{mer} \approx 1.95~\text{kHz}$, producing a stable HMNS that persists throughout the simulation duration ($\gtrsim 32~\text{ms}$). In this example, the spectrogram reveals the characteristic spectral features of a BNS merger with exceptional clarity, enabling a detailed identification of secondary peaks associated with mode couplings and other post-merger dynamics. Focusing on the spectrum within the first 2.5 ms after merger, the most readily identifiable peak corresponds to the transient dominant mode\footnote{
    Given the short duration of the strain interval, both the choice of the window function and its length prior to the Fourier transform notably influence frequency estimates. To represent this uncertainty, which adds to the intrinsic uncertainty in frequency, we assign an asymmetric uncertainty to each estimated value based on the full width at half maximum of each peak in the PSD, with left and right values representing the distance (in Hz) from the peak center to each side.} 
$f_{2,i} = 4110_{-60}^{+50}~\text{Hz}$, which subsequently evolves into the dominant mode, $f_{2} = 3670_{-180}^{+180}~\text{Hz}$~\cite{Rezzolla:2016}.
The lowest visible peak aligns with the coupling mode between the fundamental and quadrupole modes, identified as $f_1 = 1520_{-180}^{+160}~\text{Hz}$. Between $f_1$ and $f_{2,i}$, two additional short-lived peaks are observed,  
which could correspond to a rotating spiral deformation, $f_\text{spiral} = 3230_{-130}^{+100}~\text{Hz}$, and a low-frequency modulation between $f_2$ and $f_\text{spiral}$ at $2540_{-170}^{+200}~\text{Hz}$~\cite{Bauswein:2015}.
The post-merger also shows a later-emerging component, which we attribute to the coupling between the dominant mode and the quasi-radial axisymmetric mode $m = 0$, specifically $f_{2-0} = 1830_{-160}^{+180}~\text{Hz}$~\cite{Rezzolla:2016}.

\subsection{Waveform injections}

Our analysis focuses on the merger and post-merger phases, where EOS-specific information is expected to be most prominently encoded. All strains are therefore truncated from $2~\text{ms}$ before the merger, capturing a small fraction of the late inspiral to ensure merger integrity, to the end of each simulation. The only exceptions are certain GWs that exhibited a nonphysical post-merger revival, typically attributed to numerical artefacts such as resolution effects in the grid and boundary reflections. We filter out systems that experience prompt collapse into a black hole within $2~\text{ms}$ after the merger, as the available data samples for such events are considered insufficient to draw statistically significant conclusions. This selection process allows us to focus on long-lived remnants, which are more suitable for our analysis.
This approach effectively excludes from the dataset most simulations with extreme initial parameters, such as $M > 3 M_\odot$ and $q > 1.4$.

\begin{table}
    \caption{ 
        Parameters for each EOS, showing minimum, median, and maximum values. $N$ is the number of GW simulations included in the dataset, $M_\text{tot}$ denotes the total gravitational mass of the system, $q$ is the mass ratio, $e$ is the eccentricity, $\chi_A$ and $\chi_B$ are the dimensionless spin parameters of the individual stars, and $t_\text{GW}$ is the duration of the simulated GW signal from the merger onward. All values are given in geometrized units ($c = G = 1$) and solar masses ($M_\odot = 1$), except for $t_\text{GW}$, which is in milliseconds.
    }
    \label{tab:eos-parameters}
    \begin{ruledtabular}
    \begin{tabular}{lcccccccc}
        EOS & $N$ & Value & $M_\text{tot}$ & $q$ & $e$ & $\chi_A$ & $\chi_B$ & $t_\text{GW}$ \\
        \hline
        \noalign{\vskip 3pt}
        MS1b & 43 & Min & 2.500 & 1.000 & 0     & 0.187 & 0.236 & 6.6 \\
             &    & Med & 2.750 & 1.000 & 0.003 & 0.371 & 0.373 & 29.2 \\
             &    & Max & 3.400 & 2.059 & 0.156 & 0.707 & 0.764 & 69.8 \\[5pt]
        H4   & 36 & Min & 2.700 & 1.000 & 0     & 0.306 & 0.345 & 9.0 \\
             &    & Med & 2.750 & 1.000 & 0.005 & 0.495 & 0.398 & 35.5 \\
             &    & Max & 2.751 & 1.750 & 0.013 & 0.599 & 0.726 & 57.4 \\[5pt]
        DD2  & 21 & Min & 2.400 & 1.000 & 0     & 0     & 0     & 21.2 \\
             &    & Med & 2.732 & 1.092 & 0     & 0     & 0     & 24.6 \\
             &    & Max & 3.000 & 1.427 & 0     & 0     & 0     & 40.5 \\[5pt]
        BLh  & 15 & Min & 2.600 & 1.000 & 0     & 0     & 0     & 13.5 \\
             &    & Med & 2.741 & 1.177 & 0     & 0     & 0     & 38.5 \\
             &    & Max & 2.900 & 1.664 & 0     & 0     & 0     & 105.0 \\[5pt]
        SLy  & 13 & Min & 2.461 & 1.000 & 0     & 0.285 & 0.381 & 3.2 \\
             &    & Med & 2.701 & 1.000 & 0.001 & 0.460 & 0.439 & 17.9 \\
             &    & Max & 2.750 & 1.750 & 0.015 & 0.694 & 0.831 & 64.5 \\
    \end{tabular}
\end{ruledtabular}
\end{table}

Table~\ref{tab:eos-parameters} presents the number of selected GW simulations per EOS, along with the range of initial values and durations of the simulated strains. Notably, the table highlights the class imbalance, with the most populated EOS class (MS1b) having more than three times the number of simulations as the least populated class (SLy). Additionally, the ranges of $t_\text{GW}$ and spin parameters ($\chi_A$, $\chi_B$) vary significantly across EOSs, with MS1b exhibiting the longest $t_\text{GW}$ and the widest spin parameter distributions. The simulations from the \textsc{CoRe} collaboration are performed using code units and employ variable time steps. This is a common approach in NR simulations to efficiently capture the rapid dynamics near merger while saving computational resources during slower phases. To standardize the data for analysis and ensure compatibility with observational data sampled at fixed rates, we resample all waveforms at $16,384~\text{Hz}$, the usual sampling frequency of current ground-based detectors.

Each GW signal is projected as if observed from an equivalent distance of $8~\text{kpc}$, the approximate distance to the center of the Milky Way. This projection assumes an optimally oriented orbital plane, maximizing signal strength, and a geographic positioning of the detector coincident with the location of the Virgo detector in Cascina (Italy). Additionally, we set the sky location of the source to achieve optimal visibility (using \textsc{BayesWave} \cite{BayesWave:1,BayesWave:2,BayesWave:3}) and adjust the detection time to match that of GW170817, thereby simulating a realistic observational scenario.

\begin{figure}
    \centering
    \includegraphics[width=\columnwidth]{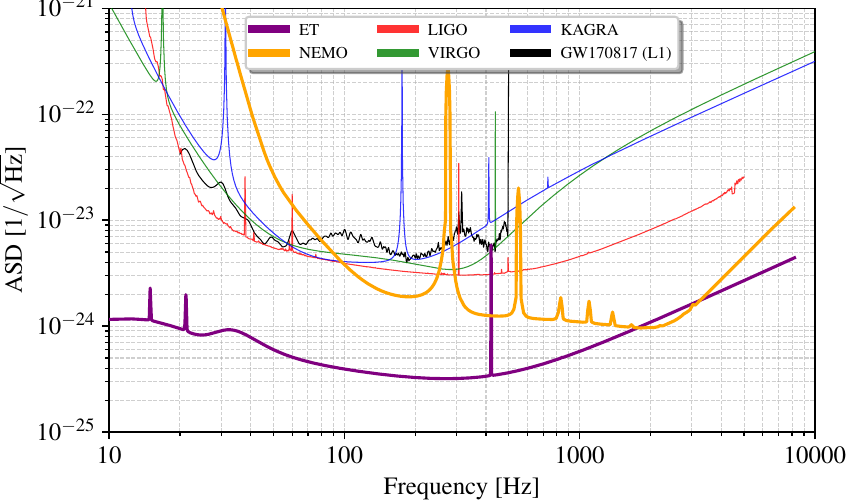}
    \caption[Design sensitivity curves for several detectors along with GW170817]{
        Comparison of the design sensitivity curves (in ASD units) for several ground-based interferometric detectors: LIGO~\cite{LIGO:2015} (red), Virgo~\cite{Acernese:2014} (green), and KAGRA~\cite{KAGRA:2020} (blue), along with the proposed next-generation detectors ET~\cite{ET:D} (purple) and NEMO~\cite{NEMO} (orange). The smoothed ASD of the GW170817 signal, as detected by the LIGO Livingston (L1) detector, is shown in black, with a frequency range between 20 Hz and 500 Hz to highlight the power excess from the merger.
    }
    \label{fig:detectors-sensitivities-gw}
\end{figure}

We next inject the GW strains into simulated background noise to replicate the expected sensitivities of the third-generation detectors ET and NEMO. This is achieved by using their proposed sensitivity curves, expressed as PSDs, which we obtained from \textsc{Bilby}~\cite{Ashton:Bilby, Bilby:NoiseCurvesNEMO}. For convenience,  Figure~\ref{fig:detectors-sensitivities-gw} compares the ASD for several detectors, including the projected sensitivities for ET and NEMO, along with the estimated ASD of the GW170817 detection by the LIGO Livingston detector. Each GW signal is injected three times, with different noise realisations for each instance. While this is not remotely enough to obtain a good statistical measure of the variance introduced by the noise in our analysis of parameters, it is intended to reduce the chance of overfitting casual correlations with it. All injections are calibrated to a SNR of 5, estimated exclusively over the merger and post-merger phases, because we are focusing on the weakest segment of the GW signal. The late inspiral segment is hence excluded from this calculation. This represents a challenging scenario for optimising the classification pipeline parameters, which we carry out under the assumption that a configuration optimised for lower SNRs will also perform well, if not better, at higher SNRs. After the injections, signals are whitened using each detector's design PSD. To exclude the low-sensitivity range of each detector's spectrum, all injections are processed through a high-pass filter, set to a cutoff frequency of 5~Hz for ET and 100~Hz for NEMO. The same spectral treatment is applied to the clean signals, as they will be used to train the dictionaries.

Our dataset therefore consists of two types of strains: the original clean GW signals and multiple injected versions of these signals, embedded in simulated background noise for ET and NEMO. For model training, we split this dataset into two subsets: $2/3$ for training and validation, and $1/3$ for the final test. To prevent data leakage, we base the split on the original clean signals. Thus, if a clean GW signal is assigned to the test set, all its injected versions are likewise assigned to the test set, ensuring no overlap between the training/validation and test data and thereby preserving the integrity of the evaluation process. Within the training and validation subset, the clean strains are used to initialize and train both the denoising and classification dictionaries, while the injected strains are dedicated to parameter optimisation and model validation.

Given the limited number of simulations available for each EOS in the training set, we employ Cross-Validation (CV) to maximize the utility of the data for both model optimisation and validation. Moreover, CV provides a less optimistic estimate of model performance during optimisation, reducing the risk of overfitting and further enhancing model generalisation on unseen data. In this work, we apply Stratified K-Fold cross-validation with $K=3$, which ensures that each fold retains a representative distribution of classes. This stratified approach mitigates the effects of class imbalance by maintaining proportionate class samples across folds. With $K=3$ and using only injections at SNR 5, each fold yields 9.33 samples for MS1b, 8.0 for H4, 4.67 for DD2, 3.33 for BLh, and 3.0 for SLy. For each fold, the distribution of samples in the training subset is truncated to integer values, with the test subset adjusted to account for any rounding discrepancies. A higher value of $K$ would reduce the number of samples available in the smallest class, causing excessive variations in the loss function due to class imbalance. Therefore, $K=3$ strikes a balance between achieving meaningful class representation within each fold and minimizing variability in loss estimates caused by class size disparities.

We also note that in order to evaluate the generalisation of our SDL model, we exclude the DD2 simulations from the main dataset. This means that the model will neither be trained nor optimised using this EOS. DD2 will be used after the final testing phase to assess the ability of our classification algorithm to associate GW signals from an unseen EOS with the class in our model that most closely matches its physical characteristics.

\section{CLASSIFICATION MODEL}\label{sec:classification-model}

The SDL-based model employed in this work builds upon the mathematical framework for denoising introduced in Llorens-Monteagudo~et~al.~(2019)~\cite{Llorens-Monteagudo:2019}, and extends it to classification using the Low-Rank Shared Dictionary Learning (\textsc{lrsdl}) model~\cite{Vu:2017}, as first applied to GW data by~\cite{Powell:2024}. As mentioned in the introduction, the classification pipeline is implemented in \textsc{clawdia}. In this section we summarise the main components relevant to the present study and refer the reader to Llorens-Monteagudo~et~al.~(2025)~\cite{CLAWDIA} for full details on \textsc{clawdia}.

The first stage of our classification model is denoising. We assume that detections in GW interferometers follow the linear degradation model
\begin{equation}
\bm{h} = \bm{u} + \bm{n},
\end{equation}
where $\bm{h}$ is the detector strain (in our case, signals injected into simulated noise), $\bm{u}$ is the underlying GW signal, and $\bm{n}$ is the detector noise. Under this model, denoising strives to recover an approximation to $\bm{u}$ by projecting $\bm{h}$ onto a representation that is meaningful only for signals of interest, and therefore unfavourable for noise.
In SDL, this representation is given by a dictionary $\bm{D} \in \mathbb{R}^{l \times a}$ whose columns (atoms) capture common patterns of the target signal population. The dimension of the dictionary is given by the product of the number of atoms $a$ and the atom length $l$. A given waveform $\bm{u}$ is approximated as a sparse linear combination of atoms,
\begin{equation}
\bm{u} \approx \bm{D} \bm{\alpha},
\end{equation}
where $\bm{\alpha}$ has only a few non-zero components. For noisy data $\bm{h}$, we obtain the coefficients by solving a minimisation problem that trades data fidelity against sparsity in $\bm{\alpha}$, known as LASSO~\cite{Tibshirani:1996},
\begin{equation}
\bm{\alpha} = \arg\min_{\bm{\alpha}} \left\{ \frac{1}{2} \| \bm{D}\bm{\alpha} - \bm{h} \|_2^2 + \lambda_\text{den} \|\bm{\alpha}\|_1 \right\}.
\end{equation}
Here, the regularisation parameter $\lambda_{\text{den}}$ controls the effective level of detail recovered. A single denoising dictionary is learned from representative clean training signals by jointly optimising $\bm{D}$ and the corresponding sparse codes over a set of segments (patches) extracted from said signals. This procedure yields atoms that encode the overall morphology of the merger and post-merger signals for all EOS models at once. In particular, we use the iterative version implemented as the \texttt{reconstruct\_iterative} method in \textsc{clawdia}. This approach enhances the stability of the hyperparameter $\lambda_{\text{den}}$, making it less sensitive to individual signal characteristics and enabling more consistent performance across diverse noise conditions.

For classification we use the \textsc{lrsdl} model, following its formulation in~\cite{Vu:2017} and its adaptation to GW data analysis in \textsc{clawdia}. The core idea is to learn a structured dictionary,
\begin{equation}
\bar{\bm{D}} = {\bm{D}_{1}, \dots, \bm{D}_{C}, \bm{D}_{0}},
\end{equation}
where each $\bm{D}_{c}\, (c=1,\cdots,C)$ contains atoms specialised to class $c$ (here, a given EOS), and $\bm{D}_{0}$ is a shared dictionary for capturing features common to multiple classes. Training is performed on clean post-merger signals labelled by EOS. The optimisation problem couples three elements: (i) sparse reconstruction of all training examples using $\bar{\bm{D}}$, (ii) Fisher discrimination dictionary learning (FDDL) constraints that encourage samples of the same class to activate similar class-specific atoms and different classes to separate in coefficient space, and (iii) a low-rank regularisation of the shared dictionary $\bm{D}_{0}$ to prevent it from absorbing class-discriminative structure. These are controlled by a totall of six hyperparameters which need to be chosen before training. Three control the dictionary dimensions, $\bm{D}_c \in \mathbb{R}^{l \times a_C}$ and $\bm{D}_0 \in \mathbb{R}^{l \times a_0}$, while the other three are the regularisation parameters related to the aforementioned optimisation elements: $\lambda_1$ (sparsity), $\lambda_2$ (Fisher term promoting both sparsity and homogeneity of the shared representation), and $\eta$ (low-rank regularisation of $\bm{D}_0$).

Given a denoised signal, \textsc{lrsdl} first computes its sparse code with respect to $\bar{\bm{D}}$. The contribution of the shared dictionary is then subtracted to isolate the class-specific content, and the signal is assigned to the class whose atoms best reconstruct the remaining content. In the present context, this means that EOS information is encoded in how well each class-specific subdictionary can represent the post-merger signal.

We optimise all hyperparameters directly for classification performance, using the macro-averaged $F_1$ score of the EOS labels as the objective function. The optimisation is carried out on low-SNR injected signals only, so that the pipeline is constrained to a rather challenging regime and is expected to generalise robustly to higher SNR. Among the denoising hyperparameters, the atom length $l$ of the denoising dictionary is treated as a key control on the frequency content and temporal locality of the learned features, and is explored through an exhaustive greedy search. Guided by previous studies, we fix the total number of dictionary features $l \cdot a$ excluding the number of atoms $a$ from the search. This choice ensures comparable computational cost for each tested value of $l$ and allows us to isolate its impact on classification accuracy.

\section{RESULTS}\label{sec:results}

\subsection{Cross-validated optimisation on the training set}\label{sec:results-optimisation}

\begin{table*}
    \caption{ 
        Cross-validation $F_1$ scores on the training set injected at SNR 5 for different dictionary lengths $l$, with the pipeline fully optimised for each length. Results are presented for both ET and NEMO detectors, with $F_1$ scores reported for each cross-validation fold and the mean $F_1$ score accompanied by its sample standard deviation.
    }
    \label{tab:eos-optimum-results-training}
    \begin{ruledtabular}
    \begin{tabular}{lcccccc}
        \multirow{2}{*}{Detector} & \multirow{2}{*}{$l$} & \multirow{2}{*}{$a$} & \multicolumn{3}{c}{Cross-Validation $F_1$ Scores} & \multirow{2}{*}{Mean $F_1$ Score} \\
        \cline{4-6}
        \noalign{\vskip 3pt}  
                                  &                      &                      & Fold 1 & Fold 2 & Fold 3 &  \\
        \hline
        \noalign{\vskip 3pt}  
        ET   & 64  & 6,400 & 0.7286 & 0.7886 & 0.8129 & $0.78 \pm 0.04$ \\
             & 128 & 3,200 & 0.7464 & 0.7782 & 0.7869 & $0.77 \pm 0.02$ \\
             & 256 & 1,600 & 0.8045 & 0.6881 & 0.8227 & $0.77 \pm 0.07$ \\[5pt]
        NEMO & 64  & 6,400 & 0.6596 & 0.6641 & 0.7661 & $0.70 \pm 0.06$ \\
             & 128 & 3,200 & 0.6769 & 0.7149 & 0.7920 & $0.73 \pm 0.06$ \\
             & 256 & 1,600 & 0.6822 & 0.7177 & 0.7611 & $0.72 \pm 0.04$ \\
    \end{tabular}
    \end{ruledtabular}
\end{table*}

\begin{table*}
    \caption{
        Optimised hyperparameters and post-training parameters for the ET and NEMO detectors. The table is divided into two sections: one for the denoising dictionary and one for the classification dictionary. The denoising dictionary parameters include $\lambda_{\text{learn}}$ and $\lambda_{\text{den}}$ (the regularisation parameters for learning and denoising, respectively), the threshold for stopping the iterative reconstruction, and the step between windows into which each signal is split. The classification dictionary parameters include $\lambda$ and $\lambda_2$ (regularisation parameters), $k$ (number of class-specific atoms), and $k_0$ (number of shared atoms).
    }
    \label{tab:eos-optimised-parameters} 
    \begin{ruledtabular}
    \begin{tabular}{l*{8}c}
        \multirow{2}{*}{Detector} & \multicolumn{4}{c}{Denoising dictionary} & \multicolumn{4}{c}{Classification dictionary} \\
        \cline{2-5} \cline{6-9} 
        \noalign{\vskip 3pt}
        & $\lambda_{\text{learn}}$ & $\lambda_{\text{den}}$ & Threshold & Step & $\lambda$ & $\lambda_2$ & $k$ & $k_0$ \\ 
        \hline
        \noalign{\vskip 3pt}  
        ET   & 0.1  & 0.5  & 0.01 & 8  & 0.01  & 0.01 & 6 & 6  \\
        NEMO & 0.1  & 0.1  & 0.01 & 16 & 0.001 & 0    & 6 & 0  \\
    \end{tabular}
    \end{ruledtabular}
\end{table*}

Before delving into the optimisation results, it is important to address key considerations regarding the pipeline's parameters. Several were fixed prior to optimisation, either due to their minimal impact on classification performance or because their behaviour was well-understood from previous studies. This deliberate approach balanced computational efficiency with the need for reliable results, allowing the optimisation process to focus on the most influential parameters.

Parameters related to the iterative denoising reconstruction, such as the threshold (set to 0.01) and maximum number of iterations (set to 1,000), were predefined based on earlier empirical tests. These values showed negligible influence on classification performance and thus did not warrant further optimisation. In contrast, the step size for signal reconstruction was optimised, as it significantly influenced classification performance, particularly for the NEMO detector.

In the training of the classification dictionary, other parameters such as the random seed and the number of training iterations were also fixed. The random seed guarantees reproducibility, while the number of iterations, set to 50, appears to be sufficient for convergence without introducing unnecessary computational overhead. Additionally, the parameter $\lambda_2$ was set to zero when $k_0 = 0$ (i.e., when only class-specific components were used), as preliminary tests showed it had negligible effects on $F_1$ scores across detectors compared to $k$. The low-rank regularisation parameter $\eta$ was similarly fixed to $\eta = 10^{-4}$ based on prior observations from~\cite{Powell:2024}, where its impact on classification was found to be minimal within an appropriate range.

Finally, although the fixed parameters were not individually optimised, their selection was guided by prior experiments and domain expertise. With this, our intention is to highlight the importance of integrating data-driven methods with expert intuition to design efficient and reliable machine learning pipelines for GW data analysis.

With this context established, Table~\ref{tab:eos-optimum-results-training} presents the cross-validation $F_1$ scores on the training set, injected at an SNR of 5, for three dictionary lengths, $l = 64$, $l = 128$, and $l = 256$. The number of atoms $a$ for each length is determined by the fixed total number of samples in the dictionary, $ C = l \cdot a = 409{,}600$, a value selected based on the available computational resources.
Each row in the table represents the best performance achieved after fully optimising the remaining pipeline parameters. Our mixed greedy and grid search optimisation approach involved testing an average of 190 parameter combinations for each $ l $. Due to the complexity of the optimisation process, full automation was not feasible, which constrained us to a narrower range of dictionary lengths.
Furthermore, $F_1$ scores are reported separately for the ET and NEMO detectors across three cross-validation folds. The last column presents the mean $F_1$ score along with its standard deviation.

Examining the ET results, we observe consistent performance across dictionary lengths, with mean $F_1$ scores ranging from 0.77 to 0.78. The length $l = 256$ yields a somewhat lower mean $F_1$ score (0.77) but with a larger standard deviation (0.07), indicating more variability across folds. In contrast, the $l = 128$ configuration has the lowest standard deviation (0.021), suggesting more stable performance, though with a slightly lower mean $F_1$ score of 0.771.

For NEMO, the mean $F_1$ score improves to a certain extent as $l$ increases from 64 to 128, reaching a peak of 0.73 with a standard deviation of 0.06. At $l = 256$, the mean score drops slightly to 0.72, with a narrower standard deviation of 0.04, suggesting stable but marginally lower performance compared to $l = 128$.

Overall, the ET scores are marginally higher than those for NEMO across all configurations, consistent with ET's higher sensitivity. The relationship between the denoising dictionary length $l$ and classification performance contrasts with simpler denoising tests, where $l$ proved to be a critical hyperparameter. Here, $l$ appears to be a stable parameter for classification, which is both due to the classification dictionary and our new iterative denoising algorithm. Given these results, we select $l = 128$ for the final test as it provides a favorable balance between mean $F_1$ score and stability across folds for both detectors.

It is important to highlight the mean standard deviation of $F_1$ scores observed across folds and dictionary lengths. Given the limited number of samples and the absence of a direct measure of $F_1$ score variance for the test set---due to statistical variability and background noise---we have chosen to use the average standard deviation across folds and lengths as an empirical threshold for determining whether a change in the measured $F_1$ score is meaningful. For the remainder of this section, we define this threshold as the ``threshold of significance'' and set it to $\Delta F_1 = 0.05$.

For completeness, in Table~\ref{tab:eos-optimised-parameters} we report the values of the main hyperparameters and post-training parameters of the pipeline optimised for the selected length of the denoising dictionary. The differences for the denoising dictionary between detectors were, in fact, minimal from the point of view of the overall classification outcome. In particular, we note that despite the apparent large difference in optimal values for $\lambda_\text{den}$, the $F_1$ score variability was negligible for each detector, so long as said values were of the same order of magnitude as the \emph{optimum} value. This further emphasizes the advantage of the iterative reconstruction method, which makes the $\lambda_\text{den}$ parameter much less dependent not only on the signal to be reconstructed, but also on the physical size of the denoising dictionary.

In contrast, the optimised hyperparameters of the classification dictionary differed significantly between detectors. The maximum number of class-specific atoms ($k$) included in the dictionary was constrained by the least populated class in the cross-validation K-Fold splits, allowing for a maximum of 6 atoms (corresponding to 6 training GW signals) per class. Based on previous work, we observed that the total number of shared atoms ($k_0$) should not exceed $k$ by much, leading us to test only a narrow range of parameter combinations.
Both detectors benefited from using the maximum number of class-specific atoms ($k = 6$), emphasizing the importance of a sufficiently rich representation for individual classes. However, the optimal configuration for shared components revealed a stark contrast. For ET, the pipeline achieved the best performance with $k_0 = 6$, indicating that a relatively large number of shared components improved classification. This suggests that ET's data contained meaningful common features across classes that the dictionary could exploit to enhance performance. The moderately low sparsity constraints ($\lambda = 0.01$, $\lambda_2 = 0.01$) further balanced feature selection without overly restricting the representation. A low value was expected, as the main sparsity constraint is already imposed through the denoising dictionary. It must be noted, however, that without shared components, the pipeline's performance was only slightly lower, which suggests that the shared dictionary might not be as relevant as initially assumed.
In fact, the optimal configuration for NEMO did not include any shared components ($k_0 = 0$), indicating the absence of shared features between classes in the injected data that the dictionary could benefit from. The lower sparsity constraint on class-specific atoms ($\lambda = 0.001$) aligns with this setting, as it allows for a broader range of class-specific features to be utilized, compensating for the challenges posed by the narrower data bandwidth.

\begin{figure}
    \begin{subfigure}[b]{\linewidth}
        \caption{ET}\label{fig:bns-confusion-matrices-train-optim:ET}
        \includegraphics[width=0.7\linewidth]{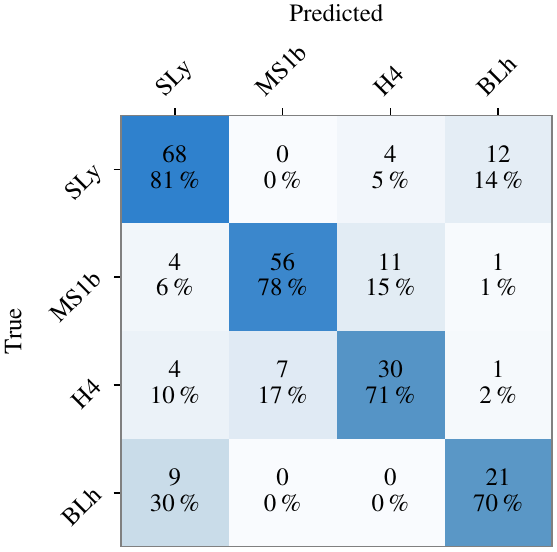}
    \end{subfigure}

    \vspace{1em}

    \begin{subfigure}[b]{\linewidth}
        \caption{NEMO}\label{fig:bns-confusion-matrices-train-optim:NEMO}
        \includegraphics[width=0.7\linewidth]{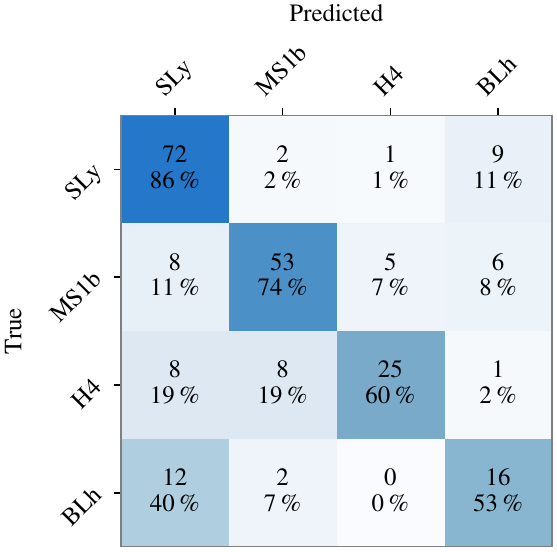}
    \end{subfigure}
    
    \caption{
        Confusion matrices for the training subset at the optimised configuration (Table~\ref{tab:eos-optimised-parameters}) for the ET (\subref{fig:bns-confusion-matrices-train-optim:ET}) and NEMO (\subref{fig:bns-confusion-matrices-train-optim:NEMO}) detectors. In each matrix, rows correspond to the true EOS, and columns to the EOS predicted by our pipeline.
    }
    \label{fig:bns-confusion-matrices-train-optim}
\end{figure}

The distribution of the classification results of the training set obtained with the optimal parameters is displayed in the confusion matrices of Figure~\ref{fig:bns-confusion-matrices-train-optim}.
%
%
The results from all CV folds have been combined into a single confusion matrix per detector by summing the fold counts. This aggregation mitigates variance from the small per-fold test sets and gives a clearer view of the overall behaviour. It smooths fold-specific fluctuations and reveals trends that single folds may obscure. The aggregated matrix does not contain independent samples, but it reduces single-fold bias and stabilises interpretation, which is especially valuable given the limited test data. However, this approach also has limitations. Aggregation obscures fold-specific insights, so strengths or weaknesses that appear in individual folds may be masked. For example, challenges with particular classes in some folds can become less apparent. Moreover, stratified splits (while balancing classes within folds) can still lead to an over-representation of frequent classes in the aggregated matrix.
For these reasons, we will only use the training results to draw conclusions on the general trend, and leave detailed descriptions and further analysis to the classification results of the test set.

In the optimised configuration, our pipeline successfully distinguishes most of the GW injections for both detectors. A systematic confusion is observed between the BLh and SLy EOS, with sufficiently high numbers to dismiss statistical variance as the source. These EOS are, in fact, the softest of our dataset and exhibit their dominant mode at the highest frequencies, where the detectors' sensitivity declines most significantly. It therefore appears reasonable that the most challenging classes to differentiate are those whose $f_2$ modes are closest to each other and at high frequencies, assuming that the key features required by the classification dictionary are predominantly concentrated in this spectral region. 
For the aforementioned reasons, drawing further conclusions from these results does not seem prudent---we reserve remaining questions for the test dataset. Nevertheless, we can still conclude that the performance demonstrated in both detectors represents a reasonable upper limit for the precision of our method at $\text{SNR} = 5$, and a promising starting point for introducing the next series of tests presented in the following sections.

\subsection{Generalisation performance} \label{sec:bns-generalisability}

We proceed to apply the optimised pipeline to the test set injected at $\text{SNR} = 5$, the same SNR as that used during cross-validation, with the objective of evaluating the generalisation ability of the optimised model. To maximize the utility of the available data, the denoising and classification dictionaries are retrained with the optimised parameters, this time using the entire training set. This approach reflects the inherent assumption of generalisation underlying the optimisation process. However, it is important to note that, given the limited size of the training dataset, a significant degree of variation in results can be anticipated.

Table~\ref{tab:eos-results-test-5snr} summarises the classification performance for both detectors, presenting precision, recall, and $F_1$ scores weighted to account for class imbalance. For the ET detector, the pipeline demonstrates moderately effective classification performance. It achieves a precision of $79\%$ and a recall (the ability to identify relevant signals) of $75\%$. The resulting $F_1$ score of $0.757$ reflects a balanced trade-off between these metrics, particularly given the low $\text{SNR}$ setting. In contrast, for the NEMO detector the pipeline exhibits weaker performance across all metrics. Precision and recall are less balanced than for ET, with only $68\%$ of relevant signals being correctly identified---around $10\%$ less than for ET. A precision of $74\%$ indicates moderate reliability of positive predictions. These differences suggest detector-specific limitations in resolving the spectral characteristics of the data.

Notably, the pipeline's performance on the test set is only slightly below that observed on the training set, which is partly expected due to the stabilising effect of cross-validation. However, it is also plausible that the final classification dictionary, retrained on the full set of original GW simulations, improved accuracy relative to the individual cross-validation folds. Some additional variability may have been introduced by the small size of the test set.

\begin{table}
    \caption{
        Performance metrics (Precision, Recall, and $F_1$ Score) for the main test, with injections performed at $\text{SNR} = 5$ in the ET and NEMO detectors.
    }
    \label{tab:eos-results-test-5snr} 
    \begin{ruledtabular}
    \begin{tabular}{lccc}
        Detector & Precision & Recall & $F_1$ Score \\
        \hline
        \noalign{\vskip 3pt}  
        ET   & 0.788 & 0.752 & 0.757 \\
        NEMO & 0.735 & 0.684 & 0.702 \\
    \end{tabular}
    \end{ruledtabular}
\end{table}

\begin{figure}
    \begin{subfigure}[b]{\linewidth}
        \caption{ET}\label{fig:bns-confusion-matrices-test-5snr:ET}
        \includegraphics[width=0.7\linewidth]{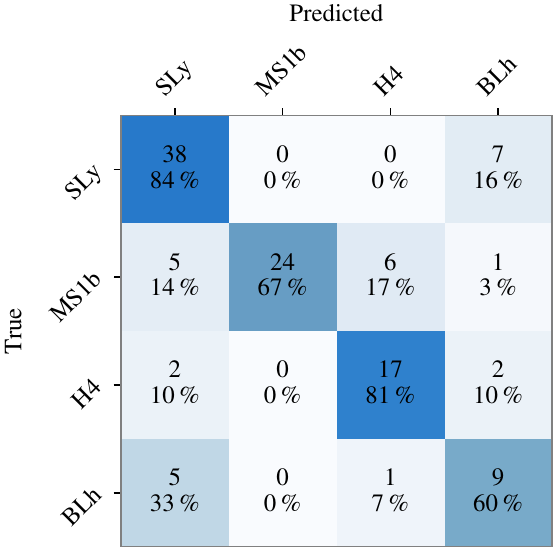}
    \end{subfigure}

    \vspace{1em}

    \begin{subfigure}[b]{\linewidth}
        \caption{NEMO}\label{fig:bns-confusion-matrices-test-5snr:NEMO}
        \includegraphics[width=0.7\linewidth]{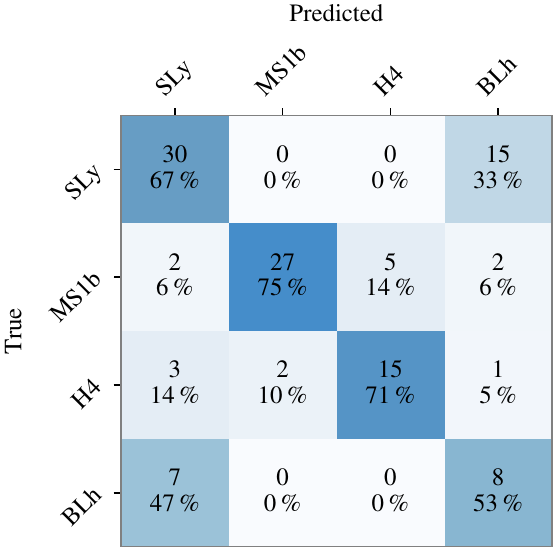}
    \end{subfigure}

    \caption{
        Confusion matrices for the test subset at the optimised configuration for the ET (\subref{fig:bns-confusion-matrices-test-5snr:ET}) and NEMO (\subref{fig:bns-confusion-matrices-test-5snr:NEMO}) detectors. In each matrix, rows correspond to the true EOS, and columns correspond to the EOS predicted by the pipeline.
    }
    \label{fig:bns-confusion-matrices-test-5snr}
\end{figure}

While the global metrics provide a high-level overview of the classification performance, they do not reveal the contributions of individual classes to the overall results or the nature of misclassifications. The natural next step is to examine the confusion matrices for both detectors, shown in Figure~\ref{fig:bns-confusion-matrices-test-5snr}, which offer a detailed view of class-specific performance. Despite the significant disparity in sensitivity between the two detectors, both share a common primary source of misclassification: a clear overlap between the SLy and BLh equations of state. In ET, 16\% of true SLy signals are misclassified as BLh, while 33\% of true BLh signals are misclassified as SLy, making this the most prominent source of confusion. In NEMO, these rates are notably higher, with 33\% of SLy signals misclassified as BLh and 47\% of BLh signals misclassified as SLy. Nevertheless, the confusion among the remaining classes is comparably balanced in both detectors. As will be shown in a subsequent noise-only test, the confusion among these other classes falls within the uncertainty introduced by noise itself. Therefore, we focus on analysing the main source of confusion.

\begin{figure}
    \centering
    \begin{subfigure}[b]{\linewidth}
        \caption{ET}\label{fig:bns-median-spectral-trends:ET}
        \includegraphics[width=\columnwidth]{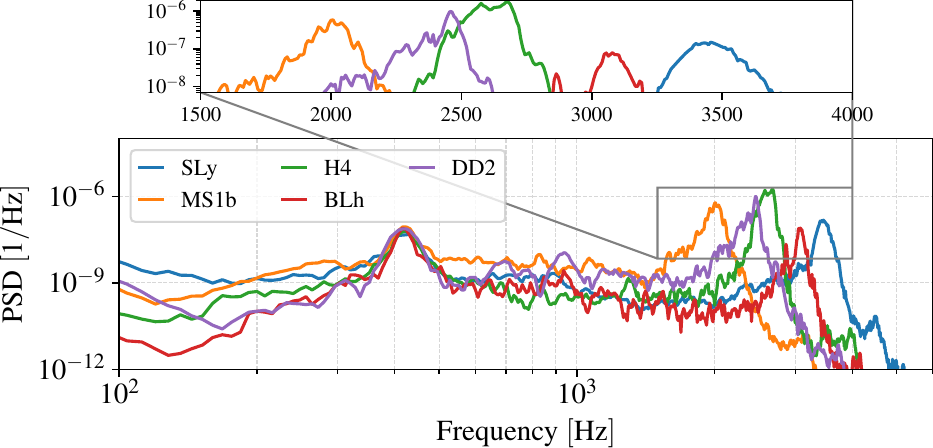}
    \end{subfigure}

    \vspace{1em}

    \begin{subfigure}[b]{\linewidth}
        \caption{NEMO}\label{fig:bns-median-spectral-trends:NEMO}
        \includegraphics[width=\columnwidth]{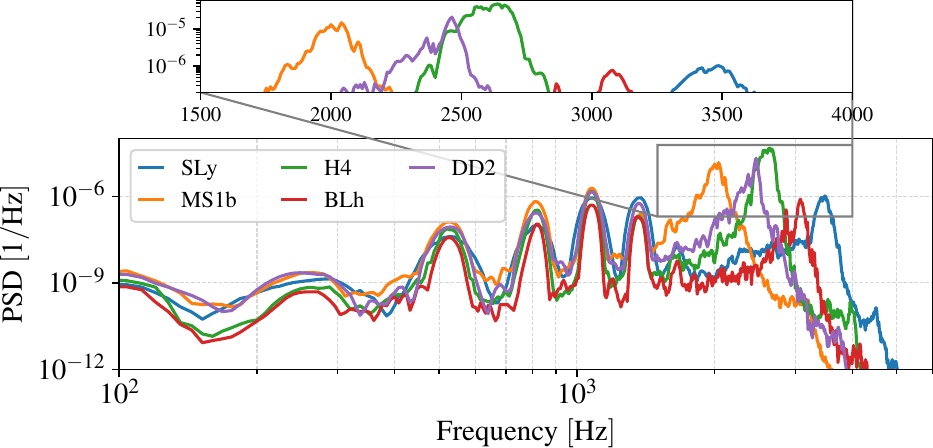}
    \end{subfigure}

    \caption{
        Average spectral distribution of the GW simulations in the test dataset for each EOS, weighted by the design sensitivity curves of the ET (\subref{fig:bns-median-spectral-trends:ET}) and NEMO (\subref{fig:bns-median-spectral-trends:NEMO}) detectors. Both axes are in logarithmic scale, representing the PSD as a function of frequency. Different colors denote each EOS, with the DD2 EOS included for reference in subsequent sections. An inset plot highlights the most prominent class-specific peaks, using a linear frequency scale.
    }
    \label{fig:bns-median-spectral-trends}
\end{figure}

To better understand the significant overlap between SLy and BLh, we analyse the spectral properties of the data, focusing on the shared spectral components across classes as well as those unique to each class. We do so computing the PSD of all whitened strains (since it emulates what the detectors would observe) using a single periodogram. The analysis concentrates specifically on the post-merger phase, considering only the data available 2.5~ms after the merger. This is motivated by the wide consensus that the dominant spectral peaks of the remnant carry the strongest EOS-dependent information and are more clearly identifiable once the transient dynamics have subsided. By contrast, the quasi-universal relations discussed in recent work mainly involve features at merger or in the very early post-merger phase~\citep{Sarin:2021,Gonzalez:2023,Topolski:2023}. 
Nevertheless, the same analysis was performed on the merger phase, spanning from -2.5 ms to 2.5 ms relative to the merger, but no notable trends or deviations were observed in that interval. This process is repeated for both detectors, with the results shown in Figure~\ref{fig:bns-median-spectral-trends}.
The top panel (\ref{fig:bns-median-spectral-trends:ET}) illustrates the spectral distribution as it would be observed by ET under ideal conditions (i.e., without background noise), while the bottom panel (\ref{fig:bns-median-spectral-trends:NEMO}) presents the corresponding distribution for NEMO.
The most prominent common features in both detectors are the peak around 420~Hz in ET and the peaks around 275~Hz, 550~Hz, 832~Hz, 1096~Hz, and 1380~Hz in NEMO. These correspond to the spectral spikes in the design sensitivity curves of their respective detectors.
In contrast, the most notable class-specific features are peaks located at the high-frequency end of the spectrum. These arise from the averaged distribution of individual peaks associated with the dominant $f_2$ mode of each GW signal, which, as explained in Section~\ref{sec:dataset}, depends on the EOS and is primarily influenced by its stiffness. Notably, the order of these peaks, from lowest to highest frequency, roughly follows the inverse order of EOS stiffness as listed in Table~\ref{tab:eos-physical-properties}.

Within the dominant mode peaks, the most notable observation is the significant difference in magnitude between the SLy (blue line) and BLh (red line) peaks compared to the remaining three EOS classes. In the ET detector, this difference spans an order of magnitude, whereas in NEMO, it increases to approximately two orders of magnitude. This pronounced drop in magnitude, caused by the sensitivity decay of both detectors at higher frequencies, likely contributes to the overlapping classification results observed between SLy and BLh.
Since the dominant peaks represent the most prominent class-specific features in the spectrum, it is reasonable to infer that they constitute the primary contribution to the class-specific components of the classification dictionary. Based on this assumption, and considering the relative proximity and reduced magnitude of the SLy and BLh peaks, it is expected that the classification dictionary faces greater challenges in distinguishing these two classes.
In the case of NEMO (where greater confusion between SLy and BLh was observed) this increased magnitude difference likely worsens precision, especially given the higher number of common frequency peaks with magnitudes comparable to the dominant peaks.

\subsection{Classification robustness under varying SNR} \label{sec:bns-snr-limit}

\begin{figure}
    \centering
    \includegraphics[width=\columnwidth]{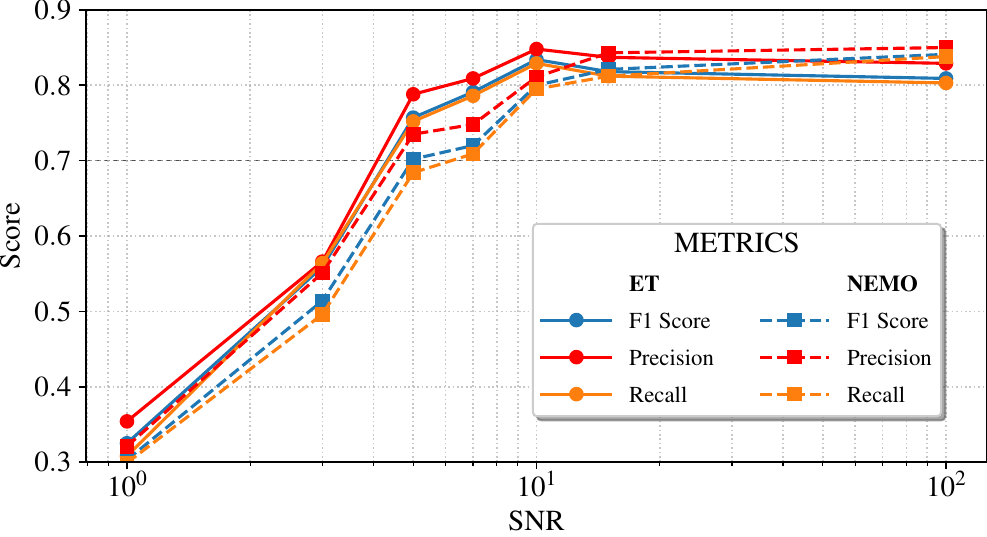}
    \caption{
        Performance trends of key metrics ($F_1$ Score, Precision, and Recall) in relation to SNR for ET (solid lines) and NEMO (dashed lines). The horizontal dashed black line at a score value of $0.7$ marks the $F_1$ Score threshold that we consider acceptable for classification performance.
    }
    \label{fig:bns-snr-limit-test-optimum-config}
\end{figure}

To assess the minimum SNR required for precise classification, we analyse the pipeline's behaviour across a range of $\text{SNR}$ levels.
Table~\ref{tab:eos-snr-limit-test-optimum-config} lists the precision, recall, and $F_1$ score for both ET and NEMO detectors as a function of $\text{SNR}$. Their trends are visually represented in Figure~\ref{fig:bns-snr-limit-test-optimum-config}.

At $\text{SNR} = 1$, the classification performance is very poor for both detectors, with $F_1$ scores of $0.325$ for ET and $0.303$ for NEMO. For four classes, these values are only marginally above what would be expected from random assignments on pure noise, indicating that the classification dictionary is effectively unable to distinguish between classes under these conditions. Consequently, we omit these injections from further analyses, as their behaviour is dominated by noise and does not provide meaningful insight into the pipeline's performance.

As the $\text{SNR}$ increases to $\text{SNR} = 3$, classification performance improves, with $F_1$ scores of $0.559$ for ET and $0.514$ for NEMO. At this level the dictionary begins to extract meaningful signal features, but substantial misclassifications remain; the classes are still not reliably separated. We retain these injections as an extreme low-$\text{SNR}$ case to examine trends under challenging conditions, but restrict specific conclusions to clearer patterns observed at higher $\text{SNR}$ levels.

At $\text{SNR} = 5$, the scenario for which the pipeline was optimised, the classification performance reaches $F_1$ scores of $0.757$ for ET and $0.702$ for NEMO. At this point the model begins to recover class-specific structure in a stable way, precision and recall are reasonably balanced, and the behaviour of the classifier is consistent across detectors.

Beyond $\text{SNR} = 5$, the performance continues to improve, with the $F_1$ score for ET increasing to $0.791$ at $\text{SNR} = 7$ and stabilising at $0.834$ at $\text{SNR} = 10$. NEMO shows similar gains, reaching $0.720$ at $\text{SNR} = 7$ and $0.800$ at $\text{SNR} = 10$. At very high $\text{SNR}$ values, such as $\text{SNR} = 15$ and $\text{SNR} = 100$, performance plateaus, with $F_1$ scores of $0.809$ for ET and $0.841$ for NEMO at $\text{SNR} = 100$. This suggests that beyond a certain threshold, the $\text{SNR}$ is no longer the primary limiting factor; instead, classification accuracy becomes constrained by intrinsic properties of the denoised dataset, the detectors' sensitivity curves, and the limited representation of the training set.

\begin{table}
    \caption{
        Performance metrics (Precision, Recall, and $F_1$ Score) for the ET and NEMO detectors across different SNR values.
    }
    \label{tab:eos-snr-limit-test-optimum-config} 
    \begin{ruledtabular}
    \begin{tabular}{r ccc ccc}
        \multirow{2}{*}{SNR} & \multicolumn{3}{c}{ET} & \multicolumn{3}{c}{NEMO} \\
        \cline{2-4} \cline{5-7} 
        \noalign{\vskip 3pt}
        & Precision & Recall & $F_1$ Score & Precision & Recall & $F_1$ Score \\ 
        \noalign{\vskip 3pt}
        \hline
        \noalign{\vskip 3pt}
        1   & 0.354 & 0.308 & 0.325 & 0.321 & 0.299 & 0.303 \\
        3   & 0.566 & 0.564 & 0.559 & 0.551 & 0.496 & 0.514 \\
        5   & 0.788 & 0.752 & 0.757 & 0.735 & 0.684 & 0.702 \\
        7   & 0.809 & 0.786 & 0.791 & 0.748 & 0.709 & 0.720 \\
        10  & 0.848 & 0.829 & 0.834 & 0.811 & 0.795 & 0.800 \\
        15  & 0.837 & 0.812 & 0.818 & 0.843 & 0.812 & 0.821 \\
        100 & 0.829 & 0.803 & 0.809 & 0.850 & 0.838 & 0.841 \\
    \end{tabular}
    \end{ruledtabular}
\end{table}

Another factor that might limit the precision at high SNR is the sparsity of the iterative reconstructions imposed by the denoising dictionary, with the regularisation parameter $\lambda_\text{den}$ set to 0.5 for ET and 0.1 for NEMO, as listed in Table~\ref{tab:eos-optimised-parameters}. It is worth noting that increasing the sparsity of the reconstruction enhances the discrimination ability of the dictionary but simultaneously limits the details of the original waveform that can be recovered. For this reason, we suspect that some class-specific features may be lost or attenuated during the denoising process. While demonstrating this effect falls outside the scope of our current study, if this hypothesis is correct, the solution would be as simple as omitting the denoising phase, as it is unnecessary at such high SNR values.

The trends in precision and recall remain closely aligned across the tested $\text{SNR}$ values. This alignment is particularly clear at $\text{SNR} = 5$, where the pipeline was tuned using the $F_1$ score, which favours a balance between both metrics. It is noteworthy that a comparable relationship between precision and recall is also observed at the lowest and highest $\text{SNR}$ values considered.

Overall, the results suggest that a practical lower bound for usable classification with this pipeline is around $\text{SNR} = 5$, which is also the value for which it was optimised. Below this level, performance is strongly affected by noise, although limited use slightly below $\text{SNR} = 5$ may still be possible under favourable conditions. Above $\text{SNR} = 5$, the pipeline exhibits increasingly stable and consistent classification behaviour across both detectors, despite not being explicitly optimised for high-SNR regimes.

\subsection{Convergence and role of the shared dictionary} \label{sec:results-convergence-shared}

A convergence test was conducted as a natural extension of the previous analysis, taking advantage of the opportunity created during the classification experiments at different $\text{SNR}$ levels. By performing classification for each $\text{SNR}$ value, it became feasible to evaluate the behaviour of the classification dictionary across varying numbers of training iterations. Significant variations in results were observed depending on the number of iterations. Furthermore, reintroducing the shared atoms ($\bm{D}_0$) into NEMO's classification dictionary revealed a more complex convergence behaviour. This was particularly interesting given that the optimal configuration did not rely on the shared component, prompting further investigation.

Although this analysis is conducted on the test set rather than the training set, it does not involve any additional parameter optimisation. As described in Section~\ref{sec:results-optimisation}, the number of training iterations for the classification dictionary was fixed during the optimisation phase and remains unchanged in subsequent tests. Instead, the goal here is to characterise the dictionary's behaviour, examining both its convergence and the impact of including or excluding the shared components. By tracking the evolution of the $F_1$ score across iterations, we can relate specific convergence patterns to the pipeline's overall performance. The convergence study uses independent noise realisations for each injection, which introduces additional variability; a more controlled analysis would reuse the same noise realisation across all injections to isolate the effect of dictionary parameters. For this reason, the present results should be regarded as preliminary, but they already provide useful qualitative insight.

\begin{figure}
    \centering
    \includegraphics[width=\columnwidth]{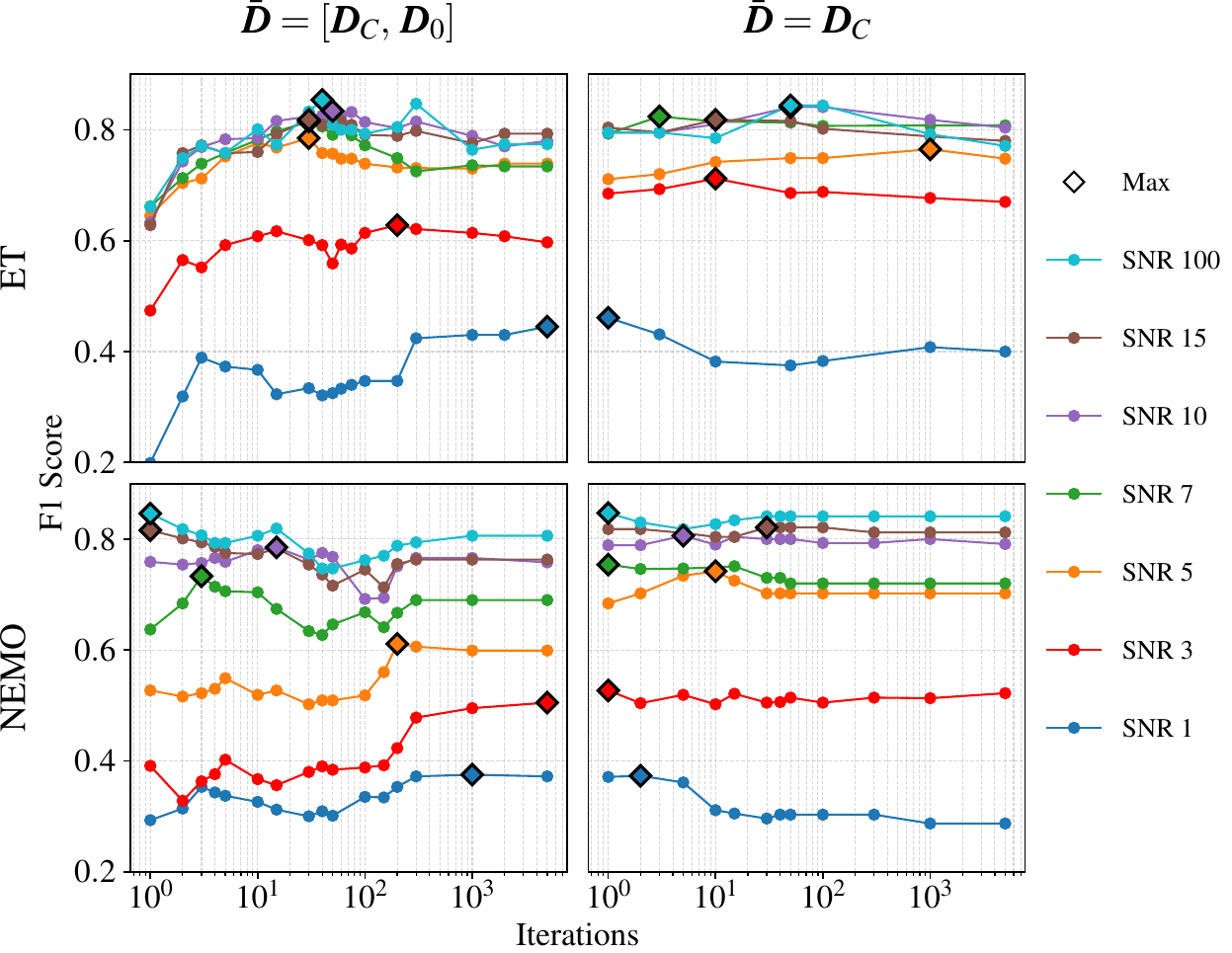}
    \caption{
        Evolution of the $F_1$ score as a function of the number of training iterations (logarithmic scale) for the ET and NEMO detectors, comparing dictionary configurations with and without the shared component $\bm{D}_0$. For each configuration, the lines show the $F_1$ score at fixed $\text{SNR}$ values as training proceeds, with the global maximum along each line highlighted by a diamond-shaped marker.
    }
    \label{fig:bns-convergence-comparison}
\end{figure}

Figure~\ref{fig:bns-convergence-comparison} presents the $F_1$ score as a function of the number of training iterations for four configurations: ET and NEMO, each evaluated with and without the shared dictionary ($\bar{\bm{D}} = [\bm{D}_C, \bm{D}_0]$ and $\bar{\bm{D}} = \bm{D}_C$). For each configuration, the data points show the evolution of the $F_1$ score at fixed $\text{SNR}$ values, with points at successive iterations joined by straight line segments; the global maximum along each line is marked by a diamond-shaped marker. This allows us to compare how quickly and how stably each dictionary configuration converges, and to visualise the differences between detectors. As in the previous analysis, the NEMO run identified as an outlier is omitted from the figure and from the conclusions.

We begin by focusing on the two scenarios that exclude the shared dictionary ($\bar{\bm{D}} = \bm{D}_C$), which simplifies the analysis.
For both detectors, the $F_1$ score remains approximately constant with respect to the number of iterations, likely due to the limited number of atoms ($k$ times the number of classes, 24 atoms of 2048 samples).
For $\text{SNR} \geq 5$, the difference in performance between ET and NEMO is not as pronounced as in the optimum configuration tested earlier and is negligible according to our threshold of significance, $\Delta \text{$F_1$} = 0.05$ (defined in Section~\ref{sec:results-optimisation}).
In Section~\ref{sec:bns-generalisability}, we hypothesized that the dominant modes constitute the primary contribution to the class-specific components. The highest frequency modes were observed to be particularly screened by noise in NEMO, which we related to the reduced performance compared to ET. This effect naturally strengthens at low $\text{SNR}$, which would explain the different scores at $\text{SNR} = 3$.

Let us now switch attention to the scenarios that include the shared dictionary ($\bar{\bm{D}} = [\bm{D}_C, \bm{D}_0]$).
In ET, for all $\text{SNR}$ values, the $F_1$ score initially increases, reaching a local maximum before stalling into a plateau. The consistency of this behaviour across all $\text{SNR}$ values indicates that, unlike the class-specific dictionary, the shared dictionary requires a minimum number of training iterations to adapt to the data, likely due to the random initialisation of shared atoms. On the one hand, at $\text{SNR} > 5$, with sufficient iterations, the best values for all trends are comparable to those without shared components. On the other hand, results for injections at $\text{SNR} = 3$ are consistently worse across all training iterations. The pipeline only appears to benefit marginally from the shared components at $\text{SNR} = 5$, the same value it was optimised for---a pattern already observed during the optimisation phase in Section~\ref{sec:results-optimisation}. And even this benefit is negligible by our significance threshold standard. Overall, these observations align with our earlier conclusion that the data does not seem to present enough common components across multiple classes. The adaptation of the shared dictionary primarily mitigates its negative impact rather than contributing with meaningful improvements to classification performance.

For NEMO, trends display slight fluctuations that barely surpass the significance threshold when noise is negligible ($\text{SNR} > 7$), resulting in a near-constant performance trend. At low $\text{SNR}$, however, trends more closely resemble those described for ET: the dictionary takes some training iterations to converge to a plateau.
To explain this behaviour, we refer to the previous spectral analysis of our data classes projected to NEMO's sensitivity in Section~\ref{sec:bns-generalisability}. In Figure~\ref{fig:bns-median-spectral-trends:NEMO}, we showed that NEMO injections display several spectral peaks common to all classes, with the intensity of these peaks varying from class to class. These class-specific features are amplified by NEMO's spectral peaks, compensating for the detector's reduced sensitivity at the highest frequencies where dominant modes lie. At high $\text{SNR}$, classes are sufficiently well-differentiated that even when shared components are enforced (which we have concluded fail to capture meaningful shared information), the classification performance is not significantly impacted.
When noise becomes dominant (red and orange lines, bottom left panel of Figure~\ref{fig:bns-convergence-comparison}), class-specific features become less prominent, and noise introduces statistical artefacts that manifest as common-to-all-classes features. These reflect inherent noise patterns rather than meaningful shared components. In this case, the shared dictionary adapts to represent these noise-driven features. While this adaptation mitigates the negative effects of including shared components, it does not improve classification performance, since these features provide no per-class information or meaningful correlations.

Overall, despite the limited statistical representation of both the signal population (simulations) and the noise (realisations)---which prevents us from drawing more detailed conclusions---we observe that for our sparse population of GW signals there is no benefit from including shared components in the classification dictionary. Regarding convergence, the number of iterations is not critical when using only class-specific components. When shared components are included, only a few iterations are required for the dictionary to stabilise. These observations apply specifically to our setup; in scenarios where shared components provide meaningful information, we would expect the number of training iterations required for convergence to increase.


\subsection{Intrinsic class imbalance}

To identify potential intrinsic biases or imbalances introduced by the classification dictionary itself, we conducted a noise-only test by injecting the same signals used in prior analyses into detector noise at $\text{SNR} = 0$, effectively nullifying the GW signal while preserving the noise realisations, and then analysed how the pipeline classified the resulting noise-only inputs.

\begin{figure}
    \begin{subfigure}[b]{\linewidth}
        \caption{ET}\label{fig:bns-intrinsic-class-imbalance:ET}
        \includegraphics[width=0.7\linewidth]{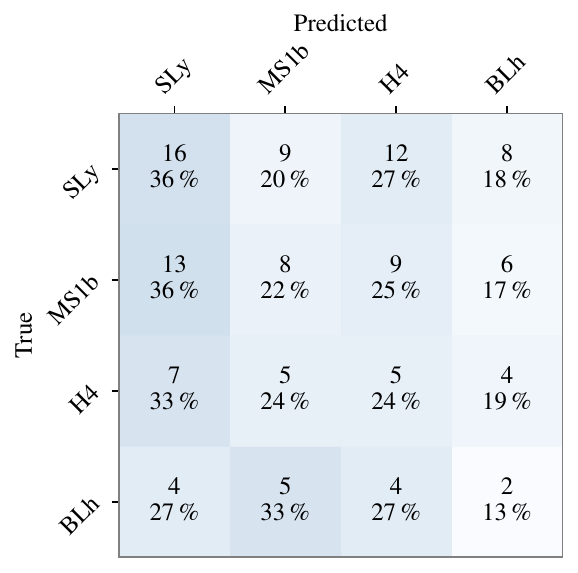}
    \end{subfigure}

    \vspace{1em}

    \begin{subfigure}[b]{\linewidth}
        \caption{NEMO}\label{fig:bns-intrinsic-class-imbalance:NEMO}
        \includegraphics[width=0.7\linewidth]{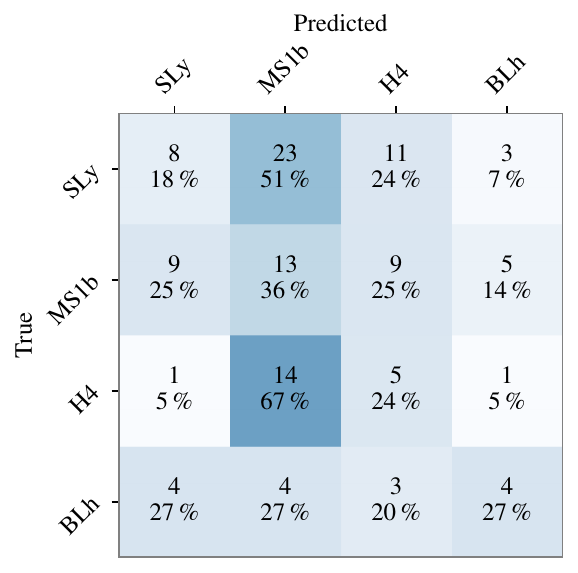}
    \end{subfigure}

    \caption{
        Confusion matrices for the classification of noise-only samples at $\text{SNR} = 0$ for ET (\subref{fig:bns-confusion-matrices-test-5snr:ET}) and NEMO (\subref{fig:bns-confusion-matrices-test-5snr:NEMO}).
    }
    \label{fig:bns-intrinsic-class-imbalance}
\end{figure}

Figure~\ref{fig:bns-intrinsic-class-imbalance} presents the confusion matrices for the noise-only test. For ET, the pipeline shows a relatively balanced classification, with no class significantly dominating the misclassifications. In contrast, the NEMO confusion matrix reveals a notable imbalance, with most noise samples being assigned to the MS1b class. This disparity suggests that the classification dictionary for NEMO, when signals are dominated by noise, exhibits a greater susceptibility to class-specific bias compared to that of ET, indicating that the bias itself is influenced by detector-specific characteristics.

\begin{figure}
    \centering
    \includegraphics[width=\columnwidth]{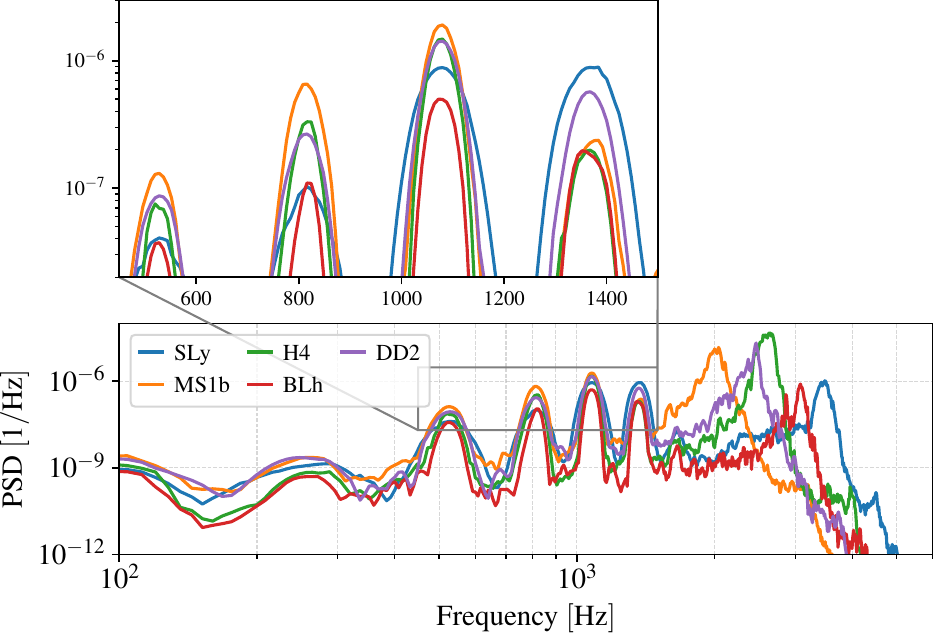}
    \caption{
        Average spectral distribution of GW signals in the test dataset for each EOS, weighted by the NEMO detector's sensitivity curve. Both axes are in logarithmic scale, representing the PSD as a function of frequency. Different colors denote each EOS, with the DD2 EOS also included for reference. The inset highlights the common spectral peaks in a linear frequency scale.
    }
    \label{fig:bns-median-spectral-trends-nemo-zoom-peaks}
\end{figure}

The dominance of MS1b in NEMO's noise-only classifications can be explained by how its class-specific dictionary components align with the detector sensitivity curve, through two related effects. First, MS1b's most prominent components lie closer to NEMO's lowest-sensitivity region than the prominent components of the other EOSs, as shown in Figure~\ref{fig:bns-median-spectral-trends:NEMO}. Second, MS1b contributes slightly more power to three of the four common peaks---thereby amplifying their spectral weight---than the rest of the EOSs. This is clearer in Figure~\ref{fig:bns-median-spectral-trends-nemo-zoom-peaks}, which shows the same average spectral distribution of GW signals for each EOS, with the inset highlighting the common peaks. Under noise-only conditions, the class with the strongest overlap in these high-power regions tends to absorb most misclassifications, which explains the observed imbalance.

Interestingly, the row-wise imbalance (true values) must arise solely from random variance in the noise realisations, as the true labels contain no inherent bias by construction. This contrasts with the column-wise imbalance (predicted values), which reflects the classification dictionary's susceptibility to detector-specific biases. From the observed standard deviation in MS1b's row, we estimate the variance expected from noise to be approximately 15\% of the true values. Extrapolating this maximum observed variance, we define a significance threshold for the predicted values (column-wise), below which any variations can be attributed to noise fluctuations rather than a true class imbalance. We conclude that the apparent class imbalance attributed to the classification dictionary may be less significant than it initially appeared.

\begin{figure}
    \begin{subfigure}[b]{\linewidth}
        \caption{ET}\label{fig:bns-intrinsic-class-imbalance-variation:ET}
        \includegraphics[width=0.7\linewidth]{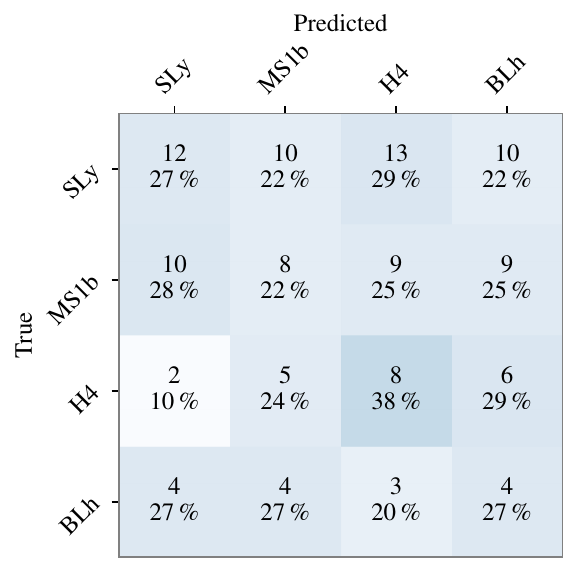}
    \end{subfigure}

    \vspace{1em}

    \begin{subfigure}[b]{\linewidth}
        \caption{NEMO}\label{fig:bns-intrinsic-class-imbalance-variation:NEMO}
        \includegraphics[width=0.7\linewidth]{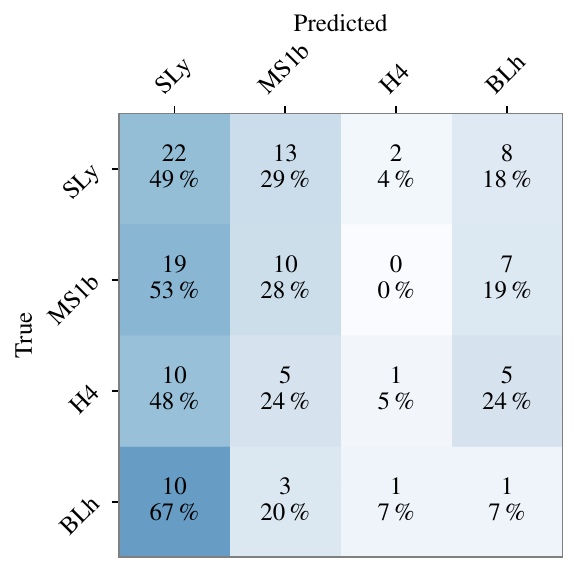}
    \end{subfigure}

    \caption{
        Confusion matrices for the classification of noise-only samples at $\text{SNR} = 0$ after modifying the shared dictionary configuration: excluding shared components in ET (\subref{fig:bns-intrinsic-class-imbalance-variation:ET}) and including shared components in NEMO (\subref{fig:bns-intrinsic-class-imbalance-variation:NEMO}).
    }
    \label{fig:bns-intrinsic-class-imbalance-variation}
\end{figure}

To further investigate the impact of the shared dictionary component ($\bm{D}_0$), we repeated the noise-only test with modified configurations: we removed shared components from ET and introduced them into NEMO. To ensure consistency without requiring additional parameter optimisation, we swapped the hyperparameters of $\bm{D}_0$ between detectors. Figure~\ref{fig:bns-intrinsic-class-imbalance-variation} presents the resulting confusion matrices. While the removal of shared components in ET produced no significant change in class balance, introducing shared components in NEMO caused the predominant class to shift from MS1b to SLy.

This shift can be understood by revisiting the spectral properties of MS1b and SLy. As shown in Figure~\ref{fig:bns-median-spectral-trends:NEMO}, MS1b dominates in most of the common spectral peaks, whereas SLy dominates only at the highest frequency common peak. When shared components are introduced, $\bm{D}_0$ competes with class-specific components to capture relevant patterns common across classes. In noise-dominated conditions, however, these common features primarily reflect noise artefacts rather than meaningful shared information. Since the main noise-driven shared features align with the intermediate spectral peaks of NEMO's sensitivity curve, the reconstruction capability of MS1b's class-specific components for reproducing these peaks diminishes. This effect applies to all classes, as these intermediate peaks are shared across the entire dataset and absorbed by the shared dictionary.

We propose that SLy emerges as the predominant class due to a combination of factors informed by prior findings, though this explanation should be taken as an educated guess rather than a definitive conclusion from the observed results. First, the shared dictionary prioritizes capturing intermediate peaks where noise artefacts dominate, leaving higher frequency regions---where SLy exhibits its strongest contributions---under-represented. Consequently, SLy’s high-frequency features remain less affected by noise-driven artefacts, allowing them to exert greater influence in the classification process. Second, the common spectral peaks in our whitened GW signals are broader than those in NEMO's sensitivity curve, primarily due to the finite resolution of the whitening process and the smoothing effect of the Hann window. This broadening effect is most pronounced at high frequencies, where the widest common peak emerges. In this scenario, SLy provides the most significant contribution to this peak, which is likely why it dominates in noise-only classifications. Third, as observed in our previous study on synthetic glitches~\citep{Llorens-Monteagudo:2019}, high-frequency oscillations are more capable of reproducing complex patterns than low-frequency oscillations when forced to do so. In~\citep{Llorens-Monteagudo:2019} the denoising dictionary successfully reconstructed Ring-Down glitches using Gaussian glitches, which are much shorter than other glitch classes. Similarly, the GW signals of the SLy EOS have their dominant mode at the highest frequencies among all classes. This combination of factors likely enables SLy's high-frequency features to compensate for the incomplete representation provided by the shared dictionary, allowing it to emerge as the predominant class in noise-only classifications.


\subsection{Classification of an unseen EOS (DD2)}

\begin{figure}
    \centering
    \begin{subfigure}[b]{\linewidth}
        \caption{ET}\label{fig:bns-classification-DD2:ET}
        \includegraphics[width=\linewidth]{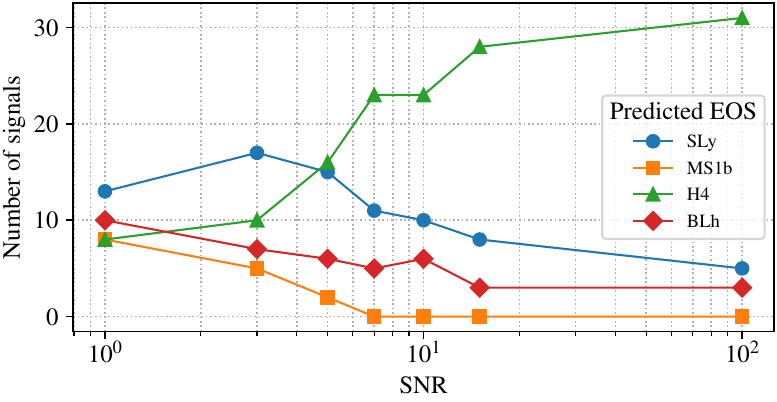}
    \end{subfigure}

    \vspace{0.5em}

    \begin{subfigure}[b]{\linewidth}
        \caption{NEMO}\label{fig:bns-classification-DD2:NEMO}
        \includegraphics[width=\linewidth]{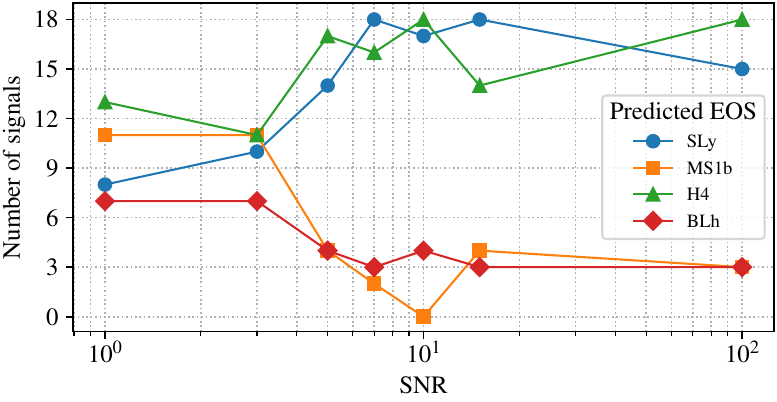}
    \end{subfigure}
    
    \caption{
        Classification results of GW signals generated with the unseen DD2 EOS, forced into the known EOS classes, for both ET (\subref{fig:bns-classification-DD2:ET}) and NEMO (\subref{fig:bns-classification-DD2:NEMO}) detectors. Each plot shows the number of signals classified into each EOS class as a function of SNR.
    }
    \label{fig:bns-classification-DD2}
\end{figure}

To evaluate the ability of our pipeline to classify GW signals from an unseen EOS (that is, an EOS whose simulated signals were excluded from both training and hyperparameter optimisation) we forced the classifier to assign DD2 signals to one of the known EOS classes. The purpose of this test is twofold: first, to determine whether the pipeline can relate waveforms from the unseen EOS to the most similar classes in our set, potentially providing physical insights based on shared characteristics between EOS; and second, to assess the SNR at which the pipeline can distinguish meaningful features. Results for both detectors are shown in Figure~\ref{fig:bns-classification-DD2}, with the number of signals classified into each EOS class plotted as a function of SNR.

For the ET detector, at low SNR values the DD2 signals are homogeneously distributed across classes. This behaviour is consistent with the pipeline's response to noise-only signals, as discussed in the previous section. As the SNR increases, however, the H4 class quickly dominates the predictions. This outcome aligns with the argument presented in Section~\ref{sec:bns-generalisability}, where the primary class-specific feature used by the classification dictionary is the spectral peak of the dominant mode. Both DD2 and H4 exhibit dominant modes that overlap significantly in the spectrum (Figure~\ref{fig:bns-median-spectral-trends}). The pipeline identifies this correlation starting at approximately $\text{SNR} = 7$, consistent with the SNR threshold established in Section~\ref{sec:bns-snr-limit} for reliable classification.

For the NEMO detector, the classification at the lowest SNR values shows a distribution similar to the noise-driven bias observed in the previous section, with MS1b being the dominant class. However, H4 already shows a significant contribution, which grows steadily with increasing SNR. Unexpectedly, SLy also emerges as a major contributor, with a classification rate comparable to H4 at higher SNR values. This divergence prompts further investigation using spectral analysis.

We refer again to Figure~\ref{fig:bns-median-spectral-trends-nemo-zoom-peaks}, which highlights the common peaks in the NEMO sensitivity curve. While the dominant modes in ET are well-separated for different EOS classes, NEMO's common spectral peaks play a larger role due to their relative prominence in the detector's sensitivity range. In the inset, the DD2 spectrum (purple line) overlaps with H4 (green line) in most peaks, except for the peak at approximately 1380~Hz, where it is closer to SLy (blue line). This overlap may explain why the NEMO pipeline artificially amplifies a correlation between DD2 and SLy, creating an unintended classification bias.

To summarise, our pipeline successfully identifies a reasonable correlation between the unseen DD2 EOS and H4 for the ET detector when the SNR exceeds 7. However, for the NEMO detector, the interaction between the pipeline and the detector's sensitivity introduces an artificial magnification of spectral components, resulting in an unexpected correlation between DD2 and SLy. These findings underscore the importance of accounting for detector-specific characteristics in pipeline design to avoid biases caused by amplified spectral features, particularly in low-SNR scenarios.

\section{DISCUSSION}\label{sec:discussion}

We have employed \textsc{clawdia}~\cite{CLAWDIA}, a newly developed sparse dictionary learning (SDL) framework, to classify the EOS of neutron stars using information encoded in the post-merger GW signals from simulated BNS mergers publicly available in the CoRe database~\cite{CoRe_2023}. The dataset covers five EOS models that capture a wide range of neutron-star properties. Our study has focused on the features emerging in the post-merger spectra, in particular the dominant post-merger frequency of the quadrupolar $f_2$ mode. These high-frequency spectral features are expected to be observed only by third-generation GW detectors such as ET and NEMO. The results reported in this work support the viability of our SDL-based pipeline for classifying the neutron star EOS through the study of post-merger GW signals.  

Several key insights emerge from our analysis, which we briefly summarise here. First, we observe that the performance of our classification pipeline is closely tied to the spectral features of the GW signals, particularly those arising from the dominant post-merger oscillation mode. The relative proximity of these spectral peaks for certain EOS classes, especially SLy and BLh, challenges the classification process. Detector sensitivity also plays a crucial role, although not so much for the overall magnitude than for the challenge that poses characterizing the complex shape of the detector's sensitivity curve in the kilohertz band.

Second, the robustness of the pipeline across varying SNR ratios underscores its potential applicability to realistic observational scenarios. We observe that classification performance begins to stabilise for SNR values around 5. Interestingly, the denoising process---while effective at lower SNRs---may introduce unnecessary sparsity constraints at high SNRs, suggesting room for further optimisation.

Third, our examination of the shared dictionary component reveals that its inclusion does not provide substantial benefits in this particular application. For the two detectors considered, the shared dictionary often captures noise-driven artefacts rather than meaningful shared features, particularly in low-SNR scenarios. This suggests that the use of class-specific components alone is sufficient for the task at hand.

Fourth, the noise-only tests highlight the intrinsic biases introduced by the classification dictionary, particularly for NEMO. These biases stem from the interplay between the detector sensitivity and the spectral characteristics of specific EOS classes. The characterisation of class imbalance opens potential avenues for improving classification performance. One immediate application is to implement a significance threshold, ensuring that predictions within the range of expected noise-driven variations are treated with caution. Additionally, incorporating uncertainties into the predictions could offer a more nuanced interpretation of classification outcomes, particularly in low-SNR scenarios. A more ambitious approach involves modifying the classification dictionary loss function to account for the observed imbalance. By assigning weights to compensate for the disproportionate representation of certain classes, this method could mitigate the effects of noise-driven artefacts and enhance the robustness of the pipeline. While speculative, this strategy holds promise for addressing class-specific imbalances systematically.

Finally, the ability of the pipeline to associate signals from an unseen EOS (DD2) with the most similar known classes provides evidence of its generalisation performance. While ET successfully correlated DD2 with H4, reflecting their spectral similarities, NEMO exhibited an artificial bias towards SLy due to its sensitivity characteristics. This highlights the importance of considering detector-specific effects when interpreting classification results.

To the best of our knowledge, this is the first study to perform multi-class EOS classification directly from merger and post-merger BNS waveforms generated by NR simulations, in the presence of detector noise. A previous study on GW-based EOS classification by Gonçalves et al.~\cite{Goncalves:2023} focuses on the inspiral phase. In particular, they used a transformer-based model to classify EOS from noise-free inspiral signals generated with the \texttt{IMRPhenomPv2\_NRTidalv2} approximant~\cite{Dietrich:2019}, drawing component masses uniformly in the range $1\text{--}2\,M_\odot$, whereas our NR sample, although spanning a similar interval, is strongly concentrated near equal-mass binaries. A direct comparison of performance is difficult, as their analysis is restricted to the inspiral and does not include any detector-specific response, while our study focuses on the merger and post-merger regime in simulated ET and NEMO noise. Nevertheless, the qualitative behaviour in tests with an unseen EOS is similar: in both cases the classifier tends to associate the new EOS with those training EOS that are closest in the $\Lambda(M)$ diagram (see the right panel of Figure~1 in~\cite{CoRe_2023}). In our framework this proximity manifests itself through the similarity of the dominant post-merger spectral peaks, supporting the interpretation that the SDL-based classifier is learning physically meaningful EOS-dependent features rather than purely numerical patterns.

The present study is based on simulated NR waveforms injected into Gaussian noise shaped by the design sensitivity curves of ET and NEMO. In real data, additional complications will arise from non-Gaussian and non-stationary noise, as well as from overlapping signals and instrumental artefacts. However, the pipeline operates on time–frequency structures that are expected to be robust to moderate deviations from idealised noise assumptions, and its modular design allows for the inclusion of more realistic conditioning and glitch-rejection stages. A natural next step is therefore to extend the analysis to mock data that incorporate representative non-Gaussian noise transients and, ultimately, to apply the method to real interferometer data as post-merger detections become available.

\begin{acknowledgments}
This work is supported by the Spanish Agencia Estatal de Investigaci\'on (grant PID2024-159689NB-C21) funded by 
MICIU/AEI/10.13039/501100011033 and by FEDER/EU, by the Generalitat Valenciana (Prometeo grant CIPROM/2022/49), and  by  the  European Horizon  Europe  staff  exchange  (SE)  programme HORIZON-MSCA-2021-SE-01 (grant NewFunFiCO-101086251).
\end{acknowledgments}

\bibliographystyle{apsrev4-2}
\bibliography{references}

\end{document}